\documentclass[11pt]{article}
\usepackage{graphicx}
\usepackage{latexsym}
\usepackage{amsmath,amsthm,amssymb,mathrsfs,amsfonts}
\usepackage{lineno,dsfont,color}
\usepackage[margin=2cm]{geometry}
\usepackage{setspace, caption}
\usepackage{times,float}
\usepackage[hyperfigures=true,bookmarksnumbered,colorlinks,linkcolor=blue,urlcolor=blue,citecolor=blue,bookmarks,plainpages=false,pdfpagelabels,hyperfootnotes=true]{hyperref}
\usepackage[font=normalsize]{caption}
\usepackage[font=normalsize]{subcaption}
\usepackage[square, comma,sort&compress,numbers]{natbib}
\usepackage{lipsum}
\usepackage{authblk}
\usepackage{etoolbox}
\allowdisplaybreaks

\newtheorem{thm}{Theorem}[section]

\numberwithin{equation}{section}
\allowdisplaybreaks[4]
\makeatletter
\long\def\MaketitleBox{%
  \resetTitleCounters
  \def\baselinestretch{1}%
  \begin{center}%
   \def\baselinestretch{1}%
    \Large\@title\par\vskip18pt
    \normalsize\elsauthors\par\vskip10pt
    \footnotesize\itshape\elsaddress\par\vskip36pt
    \end{center}%
   }

\begin{document}
\title{Mathematical assessment of the impact of non-pharmaceutical interventions on curtailing the 2019 novel Coronavirus}
\author{Calistus N. Ngonghala$^{1}$, Enahoro Iboi$^{2}$, Steffen Eikenberry$^{2}$, Matthew Scotch$^{3}$, Chandini Raina MacIntyre$^{4, \dag}$, Matthew H. Bonds$^{5}$ and Abba B. Gumel$^{2,\ddag}$ 
\footnote{Corresponding author: Email: {\it agumel@asu.edu}}\\
{\it {\small $^1$ Department of Mathematics, University of Florida, Gainesville, FL 32611, USA.}}\\
{\it {\small $^2$  School of Mathematical and Statistical Sciences, Arizona State University, Tempe, Arizona, 85287, USA.}}\\
{\it {\small $^3$ Biodesign Institute, Arizona State University, Tempe, Arizona, 85287, USA.}}\\
{\it {\small $4$ Faculty of Medicine, The Kirby Institute, University of New South Wales, Sydney, Australia.}}\\
{\it{\small $^5$ Department of Global Health and Social Medicine, Harvard Medical School, Boston, MA 02115, USA.}}\\
{\it{\small $^{\dag}$  {\it Other affiliation:}  College of Health Solutions \& College of Public Affairs and Community Solutions, Arizona State University, Tempe, Arizona, 85287, USA.}}\\
{\it{\small $^{\ddag}$  {\it Other affiliation:} Department of Mathematics and Applied Mathematics, University of Pretoria, Pretoria 0002, South Africa.}}
}

\date{}
\maketitle
\begin{abstract}
\noindent A pandemic of a novel Coronavirus emerged in December of 2019 (COVID-19), causing devastating public health impact across the world. In the absence of a safe and effective vaccine or antivirals, strategies for controlling and mitigating the burden of the pandemic are focused on non-pharmaceutical interventions, such as social-distancing, contact-tracing, quarantine, isolation, and the use of face-masks in public. We develop a new mathematical model for assessing the population-level impact of the aforementioned control and mitigation strategies. Rigorous analysis of the model shows that the disease-free equilibrium is locally-asymptotically stable if a certain epidemiological threshold, known as the {\it reproduction number} (denoted by ${\mathscr R}_c$), is less than unity. Simulations of the model, using data relevant to COVID-19 transmission dynamics in the US state of New York and the entire US, show that the pandemic burden will peak in mid and late April, respectively. The worst-case scenario projections for cumulative mortality (based on the baseline levels of anti-COVID non-pharmaceutical interventions considered in the study) decrease dramatically by $80\%$ and $64\%$, respectively, if the strict social-distancing measures implemented are maintained until the end of May or June, 2020. The duration and timing of the relaxation or termination of the strict social-distancing measures are crucially-important in determining the future trajectory of the COVID-19 pandemic. This study shows that early termination of the strict social-distancing measures could trigger a devastating second wave with burden similar to those projected before the onset of the strict social-distancing measures were implemented. The use of efficacious face-masks (such as surgical masks, with estimated efficacy $\ge 70\%$) in public could lead to the elimination of the pandemic if at least $70\%$ of the residents of New York state use such masks in public consistently (nationwide, a compliance of at least $80\%$ will be required using such masks). The use of low efficacy masks, such as cloth masks (of estimated efficacy less than $30\%$), could also lead to significant reduction of COVID-19 burden ({\it albeit}, they are not able to lead to elimination). Combining low efficacy masks with improved levels of the other anti-COVID-19 intervention strategies can lead to the elimination of the pandemic. This study emphasizes the important role social-distancing plays in curtailing the burden of COVID-19. Increases in the adherence level of social-distancing protocols result in dramatic reduction of the burden of the pandemic, and the timely implementation of social-distancing measures in numerous states of the US may have averted a catastrophic outcome with respect to the burden of COVID-19. Using face-masks in public (including the low efficacy cloth masks) is very useful in minimizing community transmission and burden of COVID-19, provided their coverage level is high. The masks coverage needed to eliminate COVID-19 decreases if the masks-based intervention is combined with the strict social-distancing strategy. 
\end{abstract}


\section{Introduction}
The world is currently facing a pandemic of a novel coronavirus (COVID-19), which started as an outbreak of pneumonia of unknown cause in Wuhan city of China in December of 2019 \cite{WHO, WHO2020a, li2020early}. As of April 14, 2020, COVID-19 (caused by the novel SARS-CoV-2 coronavirus) has spread to over 210 countries and territories, causing about 1.9 million infections and 125,000 deaths \cite{dong2020interactive, Dong_tool, WHO_SituationReports, CDC}. The United States is now the epicenter of the COVID-19 pandemic, (recording over 613,800 confirmed cases and 26,000 deaths) within a short time, with New York state bearing the brunt of the US burden (over 203,000 confirmed cases and 10,800 deaths) \cite{dong2020interactive, Dong_tool, WHO_SituationReports, CDC}. The first documented confirmed case of the novel coronavirus in the US (a resident who had recently visited Wuhan city in China) was reported on January 20, 2020 \cite{holshue2020first}, while the first confirmed case in the state of New York was reported on March 1, 2020. Most of the COVID-19 related deaths and severe cases occur in the elderly (65 years of age and older) and people with underlying medical conditions \cite{Bryner}. Younger people and frontline healthcare workers are also at high risk of acquiring COVID-19 infection. 

\par As with two other coronaviruses that caused major outbreaks in humans in recent years (namely, the Severe Acute respiratory Syndrome and the Middle Eastern Respiratory Syndrome \cite{WHO2020a, yin2018mers}), COVID-19 is transmitted from human-to-human through direct contact with contaminated objects or surfaces and through inhalation of respiratory droplets from both symptomatic and asymptomatically-infectious humans \cite{bai2020presumed}.  There is also limited evidence that the virus can be exhaled through normal breathing \cite{usnas2020}. The incubation period of the disease ranges from $2$-$14$ days, with about $97.5\%$ of infected people developing disease symptoms within 11.5 days of infection \cite{li2020early, lauer2020incubation, lai2020severe, del2020covid}. Most infections (over 80\%) show mild or no symptoms \cite{world2020coronavirus}. Common symptoms of the disease include fever, coughing and shortness of breath for mild cases, and pneumonia for severe cases \cite{CDC, WHO, WHO2020a}. In the absence of pharmaceutical interventions (such as a safe and effective vaccine for use in humans and a COVID-19 anti-viral), efforts aimed at containing COVID-19 are focused on the implementation of non-pharmaceutical interventions, such as social-distancing, using face-masks, quarantine of suspected cases, isolation and hospitalization of confirmed cases, contact-tracing and quarantine, mass testing, etc.

\par  In particular, since the novel coronavirus is transmitted among people who come in close contact with each other, the implementation of strict social-distancing measures has been the primary tool for curbing the spread of the pandemic.  As of April 7, 2020, stringent social-distancing mechanisms (mandatory lockdowns/ stay-at-home orders) have been imposed in over $42$ states of the United States, together with Washington DC, Guam, and Puerto Rico (representing over $95\%$ of the US population; involving approximately 316 million Americans) \cite{Mervosh}. The state of New York  (the current epicenter for COVID-19) has even imposed a fine against people who fail to comply with its stringent social-distancing measures that took effect March 22, 2020. Common social-distancing measures or guidelines being employed in the US include temporary closures of schools and non-essential businesses, avoiding crowded events and mass gatherings, moving in-person meetings online, etc. The city of Wuhan lifted its 76-day strict lockdown on April 8, 2020 (this was done in a phased way, with the first relaxation of measures on February 9, 2020).

\par Contact-tracing is another major public health strategy for combating the spread of COVID-19.  Contact-tracing involves searching for, or identifying, individuals with whom the confirmed case has closely interacted within a certain time frame (e.g., two days prior to the onset of symptoms \cite{bi2020epidemiology}), interviewing, testing, and isolating or hospitalizing them if they have the disease \cite{WHOSituationReport72, Josh}.

\par The use of face-masks in public by members of the general population has historically been a common practice to try to limit or combat the spread of respiratory diseases, dating back to at least the 1918 H1N1 pandemic of influenza \cite{bootsma2007effect, morens2009historical, tognotti2009influenza, group2006nonpharmaceutical, luckingham1984mask, French}.  Face masks may have been instrumental in limiting the community spread of the 2002/2003 SARS epidemic in Asia (particularly in China, Singapore, Hong Kong and Taiwan) \cite{wu2004,lau2004}  as well as the containment of the COVID-19 pandemic in Taiwan \cite{wang2020}.  Face-masks have dual purposes.  If worn by a susceptible individual, the mask offers efficacy against the acquisition of infection.  On the other hand, if the wearer is already infected (but is asymptomatic or mildly-symptomatic and unaware he/she ill), the face-mask offers efficacy against their ability to transmit infection to susceptible individuals \cite{Aiello2010, Aiello2012, cowling2009, macintyre2009, usnas2020}.

\par Predicting the course or severity of a pandemic, such as COVID-19,  as well as the realistic assessment of proposed public health intervention strategies for  combating them in real time, is a major challenge to both the public health and the scientific community.  A number of models have  been developed and used to study COVID-19 dynamics. Ferguson {\it et al.} \cite{ferguson2020impact} used an agent-based model to investigate the effects of non-pharmaceutical interventions on human deaths from COVID-19, and in reducing burden on healthcare facilities and equipment. They projected that, in the absence of control measures, over $81\%$ of the populations of the US and Great Britain might become infected and COVID-19 may cause up to 2.2 million deaths in the US. Mizumoto and Chowell \cite{mizumoto2020transmission} used a mathematical model and incidence data to study changes in COVID-19 transmission potential as the outbreak progressed through the Diamond Princess. They obtained a higher reproduction number and noticed a substantial decrease in the effective reproduction number after improved quarantine was instituted. Hellewell {\it et al.} \cite{hellewell2020feasibility} used a stochastic model with COVID-19 data to examine the impact of contact-tracing and isolation on disease control, and suggested that for most instances COVID-19 spread can be contained in 3 months if these measures are highly effective. Using a stochastic model, Kucharski {\it et al.} \cite{kucharski2020early} examined the COVID-19 trajectory in Wuhan from January-February 2020, showing a reduction in transmission (a 1.3 reduction in the associated effective reproduction number of the model) when travel restrictions were implemented.  Consequently, there is a need to examine the combined impact of multiple non-pharmaceutical interventions applied together or in sequence.

\par The present study is based on the development of a new mathematical model for studying the transmission dynamics and control of the COVID-19 pandemic in the US (particularly in the state of New York, the epicenter of COVID-19). The model takes the form of a Kermack-McKendrick, compartmental, deterministic system of nonlinear differential equations \cite{kermack1927contribution}.  It incorporates features pertinent to COVID-19 transmission dynamics and control, such as the quarantine of suspected cases and the isolation/hospitalization of confirmed COVID-19 cases (similar to the models developed in \cite{denes2019modeling, feng2007final, feng2007epidemiological}). The model,  parameterized using available COVID-19 mortality data (more reliable than case data,  provides a realistic real-time assessment and estimate of the burden of the pandemic in the US state of New York, in addition to assessing some of the main intervention strategies being implemented in the state (in particular, quarantine, isolation, contact-tracing, social-distancing and the use of face-masks in public).
 
\section{Materials and Method}
\subsection{Formulation of Mathematical Model}\label{formulation}
\noindent We designed and analyzed a novel Kermack-McKendrick-type mathematical model for the transmission dynamics and control of COVID-19 in a population \cite{kermack1927contribution}. The model, which incorporates the main non-pharmaceutical interventions being implemented to curtail COVID-19 transmission in a community (such as social-distancing, quarantine of suspected cases, isolation of confirmed cases, contact-tracing, testing and use of face-masks in public), is formulated based on stratifying the total human population at time $t$, denoted by $N(t)$, into the mutually-exclusive compartments of non-quarantined susceptible ($S_u(t))$, quarantined susceptible ($S_q(t))$, non-quarantined exposed (i.e., newly-infected individuals who do not yet show symptoms of the disease and cannot transmit infection, $E_u(t))$, quarantined exposed ($E_q(t))$, symptomatically-infectious ($I_u(t))$, asymptomatically-infectious ($I_a(t))$, hospitalized/isolated ($I_h(t))$, intensive care patients ($I_{icu}(t))$ and recovered ($R(t))$, so that 
\noindent

$$N(t) = S_u(t) + S_q(t) + E_u(t) + E_q(t) + I_u(t)+I_a(t) + I_h(t) + I_{icu}(t) + R(t).$$  

\noindent It should be mentioned that the asymptomatically-infectious compartment ($I_a$) also includes those with mild symptoms of COVID-19. Data from the World Health Organization shows that about $80\%$ of COVID-19 confirmed cases show mild or no symptoms \cite{world2020coronavirus}, and that individuals in this category (particularly those who are in the 65$+$ age group or those  with pre-existing health conditions) can develop a mild form of pneumonia that might require self-isolation or hospitalization \cite{silverstein2020first, yang2020clinical, xu2020clinical, young2020epidemiologic}. Furthermore, some individuals in this compartment (particularly those who show no clinical symptoms \cite{tian2020characteristics}) can be detected (via testing and/or tracing and testing of the contacts of confirmed COVID-19 cases) and sent to self-isolation or hospitalization. It is worth mentioning that, although the self-isolation or hospitalization of individuals in the  $I_a$ class is associated with contact-tracing, the $I_a$ to $I_h$ transition can also result from improvement in (or scaling up of) testing as is the case in Iceland \cite{gudbjartsson2020spread}. In fact, in our study, contact tracing is very much inter-linked with testing. Contact tracing is carried out after a confirmed case is diagnosed (following testing/diagnosis of a confirmed case). Furthermore, individuals in the quarantine class ($S_q$ or $E_q$) are those who have been traced, following the positive diagnosis of someone they have had close contact with (i.e., a confirmed COVID-19 case they have been exposed to). People in quarantine can be susceptible or newly-infected (but unaware of their infection) status, and are continually monitored (tested) to determine their status. Those who test positive are moved to the $E_q$ class, and those who remain negative after the incubation period are returned to the $S_u$ class.  Thus, the process of quarantining individuals suspected of having had close contacts with an infectious individual can be interpreted as contact-tracing. In addition to contact-tracing (of contacts of confirmed cases), people can be placed in quarantine because of other factors, such as having travelled to areas with high COVID-19 transmission (e.g., New York, Italy, China, Spain, UK, etc.) Quarantine and isolation can either be at home (self-quarantine and self-isolation) or at designated healthcare facilities. Furthermore, hospitalization in the context of this study includes self-isolation at home and isolation at the hospital (hence, hospitalization and isolation will be used interchangeably in this study).

\par The model is given by the following deterministic system of nonlinear differential equations (where a dot represents differentiation with respect to time).

\begin{equation}\begin{array}{lcl}\label{model}
\dot{S}_u &=& - [(1 - p) + (1 - q)p + qp]\lambda S_u + \psi_q S_q,\ \\ \ \\
\dot{S}_q &=& (1 - p)\lambda S_u - (\theta_j\lambda + \psi_q)S_q,\ \\ \ \\
\dot{E}_u &=& (1 - q)p\lambda S_u - (\sigma_u + \alpha_u)E_u,\ \\ \ \\
\dot{E}_q &=& q p\lambda S_u + \alpha_u E_u + \theta_j\lambda S_q - \sigma_q E_q,\ \\ \ \\
\dot{I}_u &=& f_1\sigma_u E_u - (\gamma_u + \phi_u+\delta_u)I_u,\ \\ \ \\
\dot{I}_h &=& f_2\sigma_u E_u + r\sigma_q E_q + \phi_u I_u +\sigma_a I_a - (\gamma_h + \nu_h+\delta_h)I_h,\ \\ \ \\
\dot{I}_a &=& (1 - f_1 - f_2)\sigma_u E_u + (1 - r)\sigma_qE_q - (\gamma_a + \sigma_a+\delta_a)I_a,\ \\ \ \\
\dot{I}_{icu} &=& \nu_h I_h - (\gamma_{icu} + \delta_{icu})I_{icu},\ \\ \ \\
\dot{R} &=& \gamma_u I_u + \gamma_h I_h + \gamma_a I_a + \gamma_{icu} I_{icu},
\end{array}\end{equation}
\noindent where {\it force of infection} $\lambda$ is defined as:
\begin{equation*}
    \lambda = \frac{\beta(1 - \epsilon_M c_M)(I_u + \eta_a I_a + \eta_h I_h)}{N - \theta_q(E_q + I_h + I_{icu})},
\end{equation*}
\noindent where $\beta$ is the effective contact rate (i.e., contacts capable of leading to COVID-19 transmission), $0 < c_M \leq 1$ is the proportion of members of the public who wear face-masks (correctly and consistently) in public and $0 < \epsilon_M \leq 1$ is the efficacy of the face-masks (low values of $c_M$ imply limited use of face-masks by members of the public, while values of $c_M$ that are closer to unity imply widespread/universal use of face-masks in the community). Furthermore, values of $\epsilon_M$ that are closer to zero imply that the face-masks are not very effective in preventing acquisition (if worn by a susceptible human) or transmitting infection (if worn by a symptomatic or asymptomatically-infectious human), while $\epsilon_M$ close, or equal to, unity implies that the face-masks used by the members of the public are of near or perfect efficacy against the acquisition or transmission of infection. Reduction in the contact rate parameter ($\beta$) can be thought of a measure of effectiveness of strategies that limit contacts between people (to avoid community transmission), notably social (or physical) distancing.
The parameter $0 < \eta_a < 1$ measures the relative infectiousness of asymptomatically-infectious humans (in the $I_a$ class) in relation to symptomatic individuals (in the $I_u$ class). Similarly, $0 < \eta_h < 1$ is a modification parameter accounting for the relative infectiousness of hospitalized/isolated infectious humans (in the $I_h$ class) in relation to individuals in the $I_u$ class. The parameter $0 < \theta_q \leq 1$ is a measure of the effectiveness of quarantine, hospitalization/isolation and ICU admission to prevent infected quarantined, hospitalized/isolated individuals and ICU patients  from transmitting infection.  In particular, $\theta_q = 0$ implies that infected quarantined, hospitalized/isolated individuals and ICU patients mix freely with the rest of population and can transmit at the same rate as other infectious individuals.  On the other hand, $\theta_q = 1$ implies that the efficacy of quarantine, hospitalization/isolation and ICU admission (in preventing infected quarantined, hospitalized/isolated individuals and ICU patients from transmitting infection) is perfect.  In other words, $\theta_q = 1$ means infected quarantined, hospitalized/isolated infectious humans and ICU patients are no longer part of the actively-mixing population (hence, do not contribute to disease transmission).

\par In the model \eqref{model}, the parameter $0 < p \leq 1$ is the probability of infection {\it per} contact, while $q$ is the proportion of non-quarantined individuals that are infected at the time of quarantine. The parameter $\psi_q$ measures the rate at which quarantined-susceptible individuals revert to the wholly-susceptible non-quarantined class ($S_u$) at the end of the quarantine period. The modification parameter $0 \leq \theta_j \leq 1$ is a measure of the efficacy of quarantine to prevent the acquisition of infection by quarantined-susceptible individuals (during quarantine). It should be mentioned that, in the formulation of the model \eqref{model}, the quarantine rate is  defined as a function of the proportion of infectious individuals in the community (in particular, the quarantine rate of susceptible individuals is $(1-p)\lambda$, while that of newly-infected exposed individuals in the $E_q$ class is $qp\lambda$, where $\lambda$ is the force of infection).  In other words, the more the number of confirmed COVID-19 cases in a community, the more the number of residents of that community that are quarantined. The parameters $\sigma_u(\sigma_q)$ is the rate at which exposed non-quarantined (quarantined) individuals progress to the symptomatic (hospitalized) class (i.e., $1/\sigma_u$ and $1/\sigma_q$ is the intrinsic incubation period of non-quarantined and quarantined exposed individuals, respectively).  A proportion, $f_1$, of exposed individuals move to the $I_u$ class at the end of the incubation period (at the rate $f_1\sigma_u$).  Similarly, another proportion, $f_2$, moves to $I_h$ class (at the rate $f_2\sigma_u$, and the remaining proportion, $1-(f_1+f_2)$ moves to the $I_a$ class (at the rate $[1-(f_1+f_2)]\sigma_u$; noting that $f_1+f_2<1$). The parameter $\sigma_a$ represents the rate at which asymptomatically-infectious humans are detected ({\it via} contact-tracing) and hospitalized.    The parameter $\sigma_a$ represents the rate at which asymptomatically-infectious humans are detected ({\it via} contact-tracing) and hospitalized. A proportion, $r$, of exposed quarantined individuals move to the $I_h$ class at the end of the incubation period (at a rate $r\sigma_q$), while the remaining proportion, $1-r$, moves to the $I_a$ class (at a rate $(1-r)\sigma_q$). The parameter $\alpha_u$ represents the rate at which exposed non-quarantined individuals are detected (by contact-tracing) and placed in quarantine.  The parameter $\gamma_u (\gamma_a)(\gamma_h)(\gamma_{icu})$ is the recovery rate for individuals in the $I_u(I_a)(I_h)(I_{icu})$ class, while $\nu_h$ is the rate at which hospitalized individuals are placed into ICU.  Furthermore, the parameter $\delta_u(\delta_a)(\delta_h)(\delta_{icu})$ represents the COVID-induced mortality for individuals in the $I_u(I_a)(I_h)(I_{icu})$ class.   It is worth mentioning that, in the formulation of the model \eqref{model}, the community transmission rate, $\beta$, is assumed to be the same for both the symptomatically-infectious ($I_u$) and the asymptomatically-infectious ($I_a$) classes.  It may be possible that the community contact rate for asymptomatically-infectious individuals is higher than that of symptomatically-infectious individuals. This is due to the fact that the former are less sick (or are not even aware that they are infected), and may, therefore, be having more contacts, and causing more infections. The assumption for the homogeneity in the community contact rate allows for a more tractable assessment of the impact of social-distancing and other community contact reduction strategies, as well as mathematical tractability.

\par
\begin{figure}[h!]
    \centering
    \includegraphics[scale=0.36]{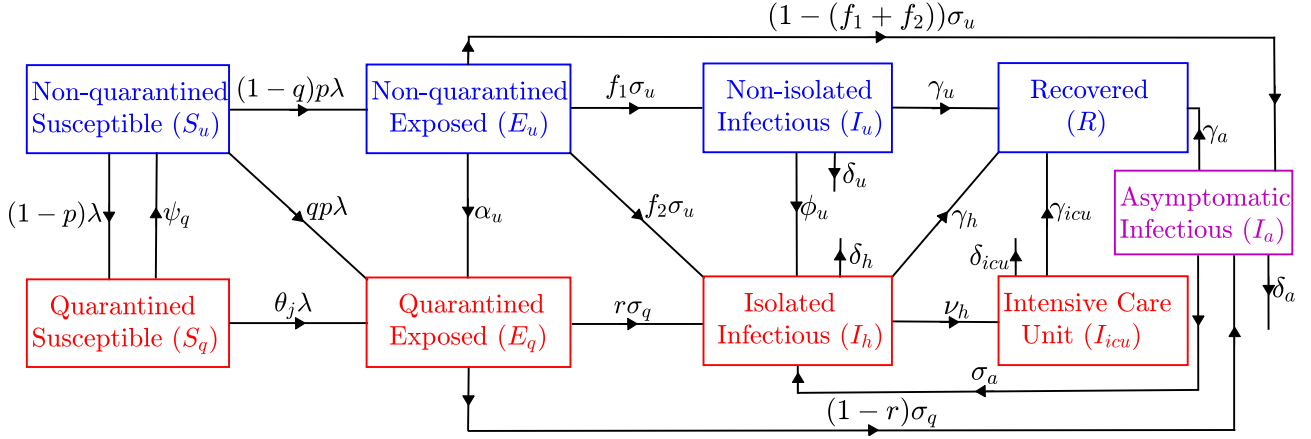}
    \caption{Flow diagram of the model \eqref{model} showing the transition of individuals between mutually-exclusive compartments based on disease status.}
    \label{fig:my_label}
\end{figure}

 The model \eqref{model} is an extension of the quarantine model for Ebola viral disease developed by Denes and Gumel \cite{denes2019modeling} (by adding epidemiological compartments for asymptomatically-infectious humans and hospitalized individuals in ICU, as well as incorporating contact-tracing of suspected cases and the use of face-masks by members of the general public). To the authors' knowledge, this may be the first deterministic model for COVID-19 that incorporates five non-pharmaceutical interventions (quarantine, isolation, contact-tracing, use of public masks, and social-distancing), in addition to allowing for the assessment of the impact of asymptomatic transmission on the trajectory and burden of COVID-19 (by adding a compartment for asymptomatic-infectious humans).
A flow diagram of the model is depicted in Figure~\ref{fig:my_label}, and the state variables and parameters of the model are tabulated in  Tables~\ref{tab:vars} and \ref{tab:params}).  \par To keep track of COVID-19 related deaths (required for calibrating our model with cumulative death data for COVID-19, and for quantifying and predicting the public health impact/burden of disease), we define the book-keeping state variable $D(t)$ to measure the number of COVID-deceased individuals. It then follows from some of the equations of the model \eqref{model} that the rate of change of the population of deceased-individuals is given by: 

\begin{equation}
    \dot{D} = \delta_u I_u + \delta_h I_h + \delta_a I_a + \delta_{icu}I_{icu}.
\end{equation}

\begin{table}[h!]
\caption{Description of state variables of the COVID-19 model.}
\begin{tabular}{|c|p{14cm}|}
\hline
{\bf State variable} & {\bf Description}\\\hline
$S_u$ & Population of non-quarantined susceptible individuals \\
$S_q$ & Population of quarantined susceptible individuals \\
$E_u$ & Population of non-quarantined exposed individuals (infected but not showing symptoms and cannot transmit infection) \\
$E_q$ & Population of quarantined exposed individuals (infected but not showing symptoms and cannot transmit infection)\\
$I_u$ & Population of symptomatically-infectious (non-hospitalized/isolated) individuals \\
$I_h$ & Population of symptomatically-infectious isolated (self/hospitalized) individuals\\
$I_a$ & Population  of asymptomatically-infectious individuals with mild or no clinical symptoms of COVID-19\\
$I_{icu}$ & Population of individuals in intensive care unit (ICU) \\
$R$ & Population of recovered individuals\\ \hline
\end{tabular}
\label{tab:vars}
\end{table}

\begin{table}[h!]
\caption{Description of the parameters of the model\eqref{model}.}
\begin{tabular}{|c|p{14cm}|}
\hline
{\bf Parameter} & {\bf Description}\\\hline
$\beta$ & Effective contact rate (a measure of social-distancing effectiveness) \\
$c_M$ & Proportion of members of public who wear masks in public (i.e., masks compliance) \\
$\epsilon_M$ & Efficacy of face-masks to prevent acquisition of infection by susceptible individuals \\
$p$  & Probability of infection {\it per} contact\\
$q$  & Proportion of infected humans quarantined at time of exposure\\
$\psi_q$ & Rate at which quarantined individuals revert to the susceptible class ($1/\psi_q$ is the average duration in quarantine).\\
$\eta_a (\eta_h)$  & Modification parameter for the assumed reduced infectiousness of asymptomatically-infectious (hospitalized/isolated) humans ($0 \leq \eta_a, \eta_h < 1$)\\
$\theta_j$  & Efficacy of quarantine to prevent acquisition of infection during quarantine $(0 \leq \theta_j < 1)$.\\
${\frac{1}{\sigma_u}} (\frac{1}{\sigma_q})$ & Incubation period for non-quarantined (quarantined) exposed individuals\\
$\sigma_a $ & Rate at which asymptomatically-infectious humans are detected ({\it via} contact-tracing) and hospitalized/isolated\\
$\alpha_u $ & Rate at which exposed non-quarantined individuals are detected ({\it via} contact-tracing) and placed in quarantine\\
$f_1 (f_2)$  & Proportion of exposed non-quarantined individuals who progress to the $I_u (I_h)$ class at the end of the incubation period ($f_1+f_2\leq 1$)\\
$1 - (f_1 + f_2)$  & Proportion of exposed non-quarantined individuals who progress to the $I_a$ class\\
$\gamma_u (\gamma_h)(\gamma_a)(\gamma_{icu})$  & Recovery rate for individuals in the $I_u (I_h) (I_a) (I_{icu})$ class\\
$\delta_u (\delta_h) (\delta_a) (\delta_{icu})$  & Disease-induced mortality rate for individuals in the $I_u (I_h) (I_a) (I_{icu})$ class\\
$\phi_u$  & Hospitalization rate of non-quarantined infectious individuals\\
$r (1-r)$  & Proportion of exposed quarantined individuals who are hospitalized (not hospitalized) at the end of the incubation period, (i.e., move to the $I_h (I_a)$ class)\\ 
$\theta_q$  & Efficacy of quarantine, hospitalization/isolation and ICU admission to prevent infected individuals in quarantine, hospital/isolation and ICU from transmitting infection ($0\leq \theta_q \leq 1)$\\ \hline
\end{tabular}
\label{tab:params}
\end{table}

\subsection{Baseline values of model parameters}
We estimated the baseline epidemiological parameters of the model from available COVID-19 data and sources from the published literature. Since the generally recommended period for quarantine of people suspected of being exposed to COVID-19 is 14 days \cite{world2020coronavirus, WHO_SituationReports}, we set the rate at which quarantined susceptible individuals revert to the non-quarantined susceptible class $(\psi_q$) to be $\psi_q = 1/14$ {\it per} day. Further, whiel some studies have estimated the incubation period for COVID-19 to range from $2$-$14$ days, with about $97.5\%$ of infected people developing disease symptoms within 11.5 days of infection \cite{lauer2020incubation, lai2020severe, del2020covid}, other studies have estimated the incubation period to be $5$-$6$ days \cite{li2020early}. We consider an average incubation period (taken from these ranges) of $5.1$ days, so that $\sigma_u=\sigma_q=1/5.1$ {\it per} day \cite{lauer2020incubation}. Similarly, we set the progression rate from the asymptomatically-infectious class ($I_a$) to the isolated/hospitalized class ($I_h)$to be $\sigma_a = 1/4$ {\it per} day. Following \cite{zou2020sars, ferguson2020impact}, we consider an infectious period of about 10 days, so that the recovery rate from COVID-19 infection ($\gamma$) is set to $\gamma_u = \gamma_{icu} = 1/10$ {\it per} day. Ferguson {\it et al.} \cite{ferguson2020impact} estimated the average time COVID-19 patients spent in hospital (for infections that do not lead to complications requiring ICU admission) to be about 8 days. Therefore, we set $\gamma_h = 1/8$ {\it per} day. Furthermore, following Ferguson {\it et al.}\cite{ferguson2020impact}, it is assumed that there is a short time lag (of about 5 days) between the onset of disease symptoms in non-quarantined humans and hospitalization.  Hence, we set the hospitalization rate ($\phi_u)$ to be $\phi_u = 1/5$ {\it per} day. Some studies have suggested that most COVID-19 infections (over $80\%$) show mild or no symptoms, about $14\%$ show severe symptoms (but without requiring ICU admission), and $6\%$ show critically-severe symptoms requiring ICU admission \cite{world2020coronavirus, anderson2020will, ECDC}. Consequently, we set $f_2 = 0.2$ and assume that half of the $80\%$ of cases that show no or mild symptoms are asymptomatic (hence, we set $f_1 = 0.4$ and $1 - (f_1 + f_2) = 0.4$). The modification parameter for the relative infectiousness of asymptomatic people ($\eta_a$) was estimated from \cite{ferguson2020impact,li2020} to be $0.5$.  Further, Li {\it et al.} \cite{li2020} estimated this parameter to be between 0.42 and 0.55. Hence, we set $\eta_a=\eta_h=0.5$.  Since about 15\% of COVID-19 patients die \cite{ferguson2020impact}, we estimated $\delta_u=\delta_h=0.015$ {\it per} day. To obtain estimates for $\delta_a$ and $\delta_{icu}$, we assume that $\delta_a=0.5\delta_u$ (so that $\delta_a= 0.0075$ {\it per} day) and $\delta_{icu}=1.5\delta_u$ (so that, $\delta_{icu}=0.0225$ {\it per} day) \cite{eikenberry2020}. The parameter for the efficacy of quarantine to prevent acquisition of infection during quarantine ($\theta_j$) is estimated to be $\theta_j=0.5$. 

\begin{figure}[H]
\centering
\includegraphics[width=0.8\textwidth]{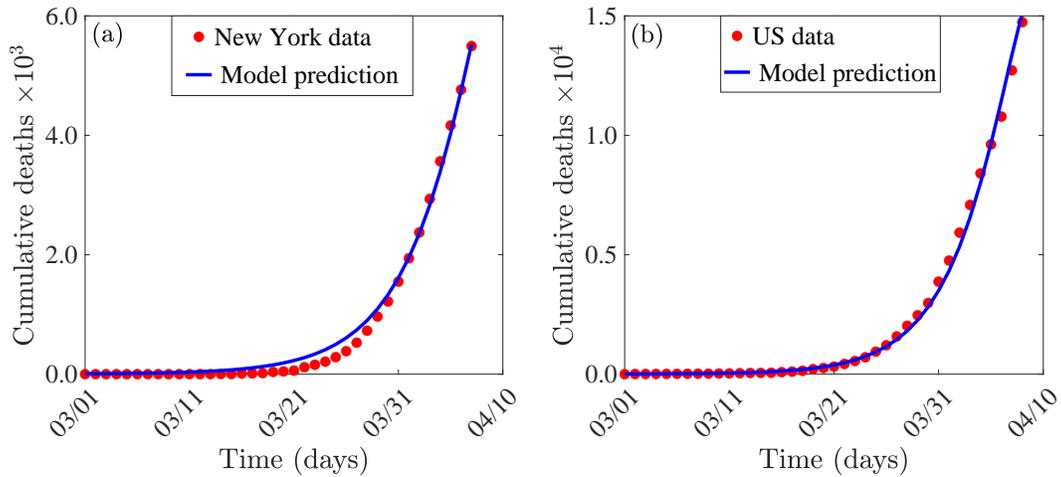}
\caption{Time series plot showing a least squares fit of system $\eqref{model}$ to New York state (a), and entire US (b), COVID-19 related death data \cite{dong2020interactive, Dong_tool, WHO_SituationReports, CDC}. The red dots represent data points, while the solid blue line represent predictions of death from system $\eqref{model}$.}
\label{ModelFit}
\end{figure}

We estimated the efficacy of face-masks ($\epsilon_M$) based on the results of a number of clinical trials. For instance, data from Driessche {\it et al.} \cite{driessche2015} shows that surgical masks reduced {\it P. aeruginosa} infected aerosols produced by coughing by over 80\% in cystic fibrosis patients.  A similar study by Stockwell {\it et al.} \cite{stockwell2018} shows that surgical masks reduced colony-forming unit (CFU) count by over 90\% (these two studies in \cite{driessche2015, stockwell2018} show that the  N95 masks (respirators) were more effective). Similarly, van der Sande {\it et al.} \cite{vandersande2008} show that home-made tea-cloth masks had an inward efficiency between 58\% and 77\% over a 3-hour duration of wear, while inward efficiency ranged 72–85\% and 98–99\% for surgical and N95-equivalent masks. Consequently, following Eikenberry {\it et al.} \cite{eikenberry2020}, we estimate inward mask efficacy to range widely between 20–80\% for cloth masks, and at least 50\% for well-made, tightly fitting masks made of optimal materials, and 70–90\% for surgical masks, and over 95\% typical for p in the range of properly worn N95 masks. Based on this, we set $\epsilon_M=0.5$.  We set the proportion of quarantined exposed individuals who are hospitalized ($r$) to be $r=0.7$.  There is no good data for the efficacy of isolation/hospitalization to prevent disease transmission by symptomatic individuals in isolation/hospital. Nonetheless, it seems plausible to assume that, at the later stage of the pandemic (such as at the present moment), the public health system capacity has been greatly improved to the extent that such transmission do not occur. Hence, we set $\theta_q=1$.  The parameter ($\nu_h$) for the rate of ICU admission is estimated to be $\nu_h=0.083$ {\it per} day (based on data from \cite{grasselli2020} it is in the range of $6\%-12\%$). The remaining parameters of the model \eqref{model}, $\beta$, $c_M$, $\sigma_a$ and $\alpha_u$, were estimated from the mortality data for the state of New York and the entire US \cite{dong2020interactive, Dong_tool, WHO_SituationReports, CDC} (based on the fitting of the state of New York data in Figure \ref{ModelFit} (a) and data for the entire US in Figure \ref{ModelFit} (b)). The estimating process involved minimizing the sum of the squares of the difference between the predictions of model \eqref{model} (cumulative deaths) and the observed COVID-19 cumulative deaths data from New York state (for the period from March 1, 2020 to April 7, 2020). In particular, the fitted values obtained using the state of New York mortality data were $\beta = 0.8648$ {\it per} day, $p = 0.8073$ {\it per} day, $c_M = 0.0546$, $\sigma_a = 0.2435$ {\it per} day and $\alpha_u = 0.1168$ {\it per} day, while the fitted values obtained using mortality data for the entire US were $\beta = 1.0966$ {\it per} day, $p = 0.7163$, $c_M = 0.1598$, $\sigma_a = 0.3090$ {\it per} day and $\alpha_u = 0.1065$ {\it per} day.

\section{Results}
\subsection{Analytical Results: Asymptotic Stability Analysis of Disease-free Equilibrium}\label{results}
 The model \eqref{model} has a line of disease-free equilibria (DFE), given by
 $$(S_u^*, S_q^*, E_u^*, E_q^*, I_u^*, I_h^*, I_a^*, I_{icu}^*, R^*) = (S_u(0), 0, 0, 0, 0, 0, 0, 0, 0),$$
 \noindent where $S_u(0)$ is the initial size of the non-quarantined susceptible individuals.  The asymptotic stability of the DFE will be analysed using the next generation operator method \cite{D-W, D-H}.  Using the notation in \cite{D-W}, it follows that the next generation operator matrices, $F$ and $V$ for the new infection terms and the transition terms, are given, respectively, by
\[
 F = 
\left[
\begin{array}{*{20}c}
 0&0&\beta_m(1 - \epsilon_Mc_M)(1-q)p &\beta_m(1 - \epsilon_Mc_M)(1-q)p \eta_h &\beta_m(1 - \epsilon_Mc_M)(1-q)p \eta_a &0&\\
 0&0&\beta_m(1 - \epsilon_Mc_M) qp &\beta_m(1 - \epsilon_Mc_M) qp \eta_h &\beta_m(1 - \epsilon_Mc_M) qp \eta_a &0&\\
 0&0&0&0&0&0&\\
 0&0&0&0&0&0&\\
  0&0&0&0&0&0&\\
 0&0&0&0&0&0&\
 \end{array}
\right],
\]
and,

\[
V =
\left[
\begin{array}{*{20}c}
 K_1&0&0&0&0&0&\\
 -\alpha_u&\sigma_q&0&0&0&0&\\
 -f_1\sigma_u&0&K_2&0&0&0&\\
 -f_2\sigma_u&-r\sigma_q&-\phi_u&K_3&-\sigma_a&0&\\
 -(1-f_1-f_2)&-(1-r)\sigma_q&0&0&K_4&0&\\
  0&0&0&-\nu_h&0&K_5&\
 \end{array}
\right],
\]
\noindent
where,  
$ K_1=\sigma_u + \alpha_u,
K_2=\gamma_u + \phi_u+\delta_u,
K_3=\gamma_h + \nu_h+\delta_h,
K_4= \gamma_a + \sigma_a+ \delta_a,
K_5= \delta_{icu} + \gamma_{icu}.$ 
\noindent It is convenient to define ${\mathcal R}_c$ by (where $\rho$ now represents the spectral radius of the next generation matrix $FV^{-1}$): 

\begin{equation}
    \mathscr{R}_c =\rho(FV^{-1}) = \mathscr{R}_{cu} + \mathscr{R}_{ca} + \mathscr{R}_{ch},\label{Rc}
\end{equation}
where, 
\begin{eqnarray*}
\mathscr{R}_{cu} &=& \frac{\beta_m(1 - \epsilon_M c_M)p B_{u}}{\displaystyle{K_1K_2K_3K_4}},~~~
\mathscr{R}_{ca} = \frac{\beta_m(1 - \epsilon_M c_M)p\eta_a B_a}{\displaystyle{K_1K_2K_3K_4}},~~~
\mathscr{R}_{ch} = \frac{\beta_m(1 - \epsilon_M c_M)p\eta_h B_h}{\displaystyle{K_1K_2K_3K_4}},\end{eqnarray*}
\noindent {\rm with},
\begin{eqnarray*}
B_{u} &=& K_3K_4(1-q)\sigma_u f_1,\\
B_a &=& K_2K_3\{(1 - r)[(1 - q)\alpha_u + K_1 q] + (1 - f_1 - f_2)( 1 - q)\sigma_u\},\\
B_h &=& [(1 - q)\alpha_u + q K_1][(1 - r)K_2 \sigma_a + r K_2K_4] + (1 - q)\sigma_u[(1 - f_1 - f_2)K_2 \sigma_a + K_4(f_1\phi_u + f_2K_2)].
\end{eqnarray*} 
\noindent The result below follows from Theorem 2 of \cite{D-W}.

\begin{thm}\label{thm1}
The disease-free equilibrium (DFE) of the  model \eqref{model} is locally-asymptotically stable if $\mathscr{R}_c < 1$, and unstable if $\mathscr{R}_c > 1$.
\end{thm}
\noindent The quantity ${\mathscr R}_c$ is the {\it control reproduction number} of the model \ref{model}. It measures the average number of new COVID-19 infections generated by an average infected individual introduced into a population where basic public health interventions (quarantine, isolation, social-distancing, testing etc.) are implemented. The quantity ${\mathcal R}_c$ is the sum of the constituent reproduction numbers associated with the number of new COVID-19 cases generated by symptomatically-infectious humans (${\mathscr R}_{cu}$), hospitalized/isolated individuals (${\mathscr R}_{ch}$) and asymptomatically-infectious humans (${\mathscr R}_{ca}$). The epidemiological implication of Theorem \ref{thm1} is that a small influx of COVID-19 cases will not generate a COVID-19 outbreak if the control reproduction number $({\mathscr R}_c)$ is less than unity.  It is worth mentioning that for Kermack-McKendrick-type mathematical models with no vital/demographic dynamics (i.e., births or natural death processes) or waning immunity (to continuously feed the susceptible class), such as the model \eqref{model}, it is instructive to compute the final size of the epidemic \cite{arino2007final, brauer2017final, brauer2019final, feng2007final, brauer2008age}. The final epidemic size relations, which are natural quantities associated with the dynamics of epidemic models (with no vital/demographic dynamics), allow for the realistic quantification of disease burden and can be used to assess the impact and effectiveness of various intervention and mitigation strategies \cite{feng2007final}.  The final size relations for the epidemic model \eqref{model} are calculated in Section \ref{FinEpicSize} below.

\subsection{Computation of Final Size of the Pandemic}\label{FinEpicSize}
\noindent In this section, the final size of the COVID-19 pandemic will be calculated. Using the notation in \cite{arino2007final}, let $x \in {\rm I\!R}_{+}^6$, $y \in {\rm I\!R}_{+}^2$, and $z \in {\rm I\!R}_{+}$ represent the sets of infected, susceptible and recovered components of the model. Thus,it follows from the model \eqref{model}, that  $x(t) = $ $(E_u(t)$, $E_q(t)$, $I_u(t)$, $I_h(t)$, $I_a(t)$, $I_{icu}(t))^T$, $y(t)=(S_u(t), S_q(t))^T$ and $z(t)=R(t)$. Further, following Arino {\it et al.} \cite{arino2007final}, let $D$ be the $m\times m$ diagonal matrix whose diagonal entries, denoted by $\sigma_i (i=1,2,\cdots, m$), are the relative susceptibilities of the corresponding susceptible class. It is convenient to define $\Pi$ to be an $n\times n$ matrix with the property that the $(i,j)$ entry represents the fraction of the $j^{{\rm th}}$ susceptible compartment that goes into the $i^{\rm th}$ infected compartment upon becoming infected. Let $b$ be an $n-$dimensional row vector of relative horizontal transmissions.  Using the notation in \cite{arino2007final}, let the infection rate, $\lambda$, of the model \eqref{model} be represented by $\beta$. That is, $\lambda=\beta(x,y,z)$. It is convenient to define the $m-$dimensional vector $\Gamma=[\Gamma_1,\Gamma_2,\cdots, \Gamma_m]=\beta b V^{-1}\Pi D$ \cite{arino2007final}. It follows, in the context of the model \eqref{model}, that
 
\begin{equation*} 
b = [0, 0, 1, \eta_h, \eta_a, 0],\,\,
\Gamma = \left[\mathcal{R}_c, \frac{\beta_m(1 - \epsilon_Mc_M) [ (1-r)(K_{{3}}\eta_{{a}}+\eta_{{h}}\sigma_{{a}})+\eta_{{h}}r
K_{{4}}] \theta_j}{K_{{3}}K_{{4}}}\right],\,\, 
D = \begin{bmatrix} 
     1 & 0 \\ 
     0 & \theta_j 
     \end{bmatrix}
\end{equation*}
\noindent
and, \begin{equation*}
   \Pi=\begin{bmatrix} 1-q& 0 \\ q&1 \\ 0&0 \\ 0&0 \\ 0&0\\0&0 \end{bmatrix}. 
\end{equation*}
\noindent
Using the above change and variables and definitions, the model \eqref{model} reduces to:

\begin{equation}\label{reducedfinalsize}
 \begin{split}
  \dot{x}&=\Pi Dy\beta(x,y,z)bx-Vx,\ \\ \ \\
  \dot{y}&=-Dy\beta(x,y,z)bx,\ \\ \ \\
  \dot{z}&=Wx,
 \end{split}  
\end{equation}
\noindent where $W$ is a $k\times n$ matrix with the property that the $(i,j)$ entry represents the rate
at which individuals of the $j^{\rm th}$ infected compartment transition into the recovered ($i^{\rm th}$ $z$) compartment
upon recovery and the matrix $V$ is as defined in Section \ref{results}.  It is worth stating that the reproduction number (${\mathscr R}_c$, of the model \eqref{model} (or, equivalently, \eqref{reducedfinalsize}), can be recovered using the definition
${\mathscr R}_c=\beta(0, y_0, z_0)bV^{-1}\Pi D y_0$ given in Theorem 2.1 of \cite{arino2007final} (it should be noted that this theorem also allows for recovering the local asymptotic stability result for the family of disease-free equilibria of the model \eqref{model}, given in Section \ref{results}). Furthermore, the results below, for the final size relations of the model \eqref{model} (or, equivalently, \eqref{reducedfinalsize}), can be established using Theorem 5.1 of \cite{arino2007final}.
\begin{thm}\label{thm2}
Consider the epidemic model \eqref{model} (or, equivalently, \eqref{reducedfinalsize}). The final size
relations are given by
\begin{eqnarray}
 \ln{\left(\frac{S_u(0)}{S_u(\infty)}\right)} &\geq& \mathcal{R}_c\frac{\left[S_u(0)-S_u(\infty)\right]}{S_u(0)} + {\frac {\theta_jp\beta_m(1 - \epsilon_M c_M) \left[ (1-r)(K_{{3}}\eta_{{a}} + \eta_{{h}}\sigma_{{a}})+\eta_{{h}}r K_{{4}} \right] \left[S_u(0)-S_q(\infty)\right]}{S_u(0)K_{{3}}K_{{4}}}}\notag\\
& +& \frac{p\beta_m(1 - \epsilon_M c_M)}{S_u(0)K_1K_2K_3K_4}\left[C_1E_u(0) + C_2E_q(0) + C_3I_u(0) + C_4I_h(0) + C_5I_a(0)\right],\notag \\
S_q(\infty) &\geq& S_q(0)\left(\frac{S_u(\infty)}{S_u(0)}\right)^{\theta_j},
\label{FinSizeRel}
\end{eqnarray}
where,
\begin{eqnarray*}
C_1 &=& \Bigg\{  \bigg\{  \left( r\alpha_{{u}}+f_{{2}}\sigma_{{u}} \right) K_{{
4}}+\sigma_{{a}} \left[ (1-r)\alpha_{{u}}+(1-f_1-f_2) \right]   \bigg\} \eta_{{h}}+
\eta_{{a}}K_{{3}} \left[ (1-r)\alpha_{{u}}+(1-f_1-f_2)  \right]  \Bigg\} K_{{2}}\\
&&+\sigma_{{u}}K_{{4}}f_{{1}} \left( \eta_{{h}}\phi_{{u}}+K_{{3}}
 \right),\\
C_2 &=& K_1K_2\left[\eta_{{a}}(1-r)K_{{3}}+(1-r)\eta_{{h}}\sigma_{{a}}+r\eta_h K_{{4}}\right],\\
C_3 &=&K_1K_4(\eta_h\phi_u+K_3) ,\\
C_4 &=& K_1K_2K_4\eta_h,\\
C_5 &=& K_1K_2(\eta_a K_3+ \eta_h \sigma_a),
\end{eqnarray*}
\end{thm}
\noindent
with the parameter groupings $K_i (i = 1, 2, 3, 4)$ as defined in Section \ref{results}. It is worth mentioning that, by setting $ E_u(0) = E_q(0) = I_u(0) = I_a(0) = I_{icu}(0) = S_q(0) = 0$, with $S_u(0) > 0$ and $I_h(0) > 0$, the final size relations, given by the inequalities in \eqref{FinSizeRel}, reduce to:

\begin{equation*}
\begin{split}
\ln{\left(\frac{S_u(0)}{S_u(\infty)}\right)}&\geq \mathcal{R}_c\frac{\left[S_u(0)-S_u(\infty)\right]}{S_u(0)}+{\frac {\theta_j\beta_m(1 - \epsilon_M c_M) \left[ (1-r)(K_{{3}}\eta_{{a}}+\eta_{{h}}\sigma_{{a}})+\eta_{{h}}r
K_{{4}} \right] \left[S_u(0)-S_q(\infty)\right]}{S_u(0)K_{{3}}K_{{4}}}}\\
&+\frac{\eta_h p\beta_m(1 - \epsilon_M c_M)}{\displaystyle{S_u(0){K}_3}}I_h(0).
\end{split}
\end{equation*}

\section{Numerical simulations}\label{NumSim}
\noindent We simulated the model \eqref{model} using the baseline parameter values tabulated in Table~\ref{ParVals} (unless otherwise stated), to assess the population-level impact of the various control and mitigation strategies against the spread of COVID-19 in the US state of New York discussed in Section \ref{formulation}. We also simulated the model using the calibrated parameters in Table~\ref{ParVals1}, together with the other estimated parameters in Table~\ref{ParVals}, to assess the population-level impact of various control measures in the entire US. It should be mentioned that in all the simulations carried out, the various non-pharmaceutical interventions are maintained at their baseline values (unless otherwise stated).

\begin{table}
\centering
\caption{Estimated and fitted parameters of the model \eqref{model} using COVID-19 mortality data for the state of New York \cite{dong2020interactive, Dong_tool, WHO_SituationReports, CDC}. With this set of parameter values, the control reproduction number (${\mathscr R}_c$) is given by $\mathscr{R}_c = 1.95$.}
\begin{tabular}{|l|l|l|}
\hline
{\bf Parameters} & Value & Source\\ \hline
$\beta$ & $0.8648$ & Fitted (calibrated using New York state COVID-19 data \cite{dong2020interactive, Dong_tool, WHO_SituationReports, CDC})\\
$c_M$ & $0.0546$ & Fitted (calibrated using New York state COVID-19 data \cite{dong2020interactive, Dong_tool, WHO_SituationReports, CDC})\\
$\epsilon_M$ & $0.5$ & Estimated from \cite{davies2013}\\
$p$ & $0.8073$ & Fitted (calibrated using New York state COVID-19 data \cite{dong2020interactive, Dong_tool, WHO_SituationReports, CDC})\\
  $q$  & $0.2$ & Assumed\\
 $\psi_q$ & $1/14$ & Estimated \cite{WHO_SituationReports, world2020coronavirus}\\
  $\eta_a $ & $0.5$ & Estimated from \cite{ferguson2020impact,li2020}\\
 $\eta_h$& $0.5$ & Estimated from \cite{ferguson2020impact,li2020}\\
$\theta_j$ & $0.5$ & Assumed\\
$\sigma_u $ & $1/5.1$ & \cite{li2020early, lauer2020incubation, lai2020severe, del2020covid}\\
$\sigma_a$ & $0.2435$  & Fitted (calibrated using New York state COVID-19 data \cite{dong2020interactive, Dong_tool, WHO_SituationReports, CDC}) \\
$\sigma_q$ & $1/5.1$ & \cite{li2020early, lauer2020incubation, lai2020severe, del2020covid}\\
$f_1 $ & $0.4$  & Estimated from \cite{world2020coronavirus, anderson2020will, ECDC}\\
$f_2$ & $0.2$ & \cite{world2020coronavirus, anderson2020will, ECDC}\\
$1 - (f_1 + f_2)$ & $0.4$ & Estimated from \cite{world2020coronavirus, anderson2020will, ECDC}\\
$\alpha_u$ & $0.1160$ & Fitted (calibrated using New York state COVID-19 data \cite{dong2020interactive, Dong_tool, WHO_SituationReports, CDC}.) \\
$\gamma_u $ & $1/10$  & \cite{zou2020sars, ferguson2020impact}\\
$\gamma_h$ & $1/8$   & \cite{ferguson2020impact}\\
$\gamma_a$ & $0.13978$ & \cite{tang2020}\\
$\gamma_{icu}$ & $1/10$ & \cite{ferguson2020impact} \\
$\delta_h$& $0.015$ & \cite{ferguson2020impact}\\
$\delta_u $& $0.015$ & \cite{ferguson2020impact}\\
$\delta_a$ & $0.0075$ & Estimated from \cite{ferguson2020impact}\\
$\delta_{icu}$ & $0.0225$ & Estimated from \cite{ferguson2020impact}\\
$\phi_u$ & $1/5$ & \cite{ferguson2020impact}\\
$r$ & $0.7$ & Assumed\\
$\nu_h$ & $0.083$ & Estimated from \cite{grasselli2020} \\
 $\theta_q$ & 1 & Assumed\\ \hline 
\end{tabular}
\label{ParVals}
\end{table}

\begin{table}
\centering
\caption{Estimated and fitted parameters of the model \eqref{model} using COVID-19 mortality data for the entire US \cite{dong2020interactive, Dong_tool, WHO_SituationReports, CDC}. Using this set of parameter values, the control reproduction number (${\mathscr R}_c$) is given by $\mathscr{R}_c = 2.07$.}
\begin{tabular}{|l|l|l|}
\hline
{\bf Parameters} & Value & Source\\ \hline
$\beta$ & $1.0966$ & Fitted (calibrated using US COVID-19 data \cite{dong2020interactive, Dong_tool, WHO_SituationReports, CDC})\\
$c_M$ & $0.1598$ & Fitted (calibrated using US COVID-19 data \cite{dong2020interactive, Dong_tool, WHO_SituationReports, CDC})\\
$p$ & $0.7163$ & Fitted (calibrated using US COVID-19 data \cite{dong2020interactive, Dong_tool, WHO_SituationReports, CDC})\\
$\sigma_a$ & $0.3090$  & Fitted (calibrated using US COVID-19 data \cite{dong2020interactive, Dong_tool, WHO_SituationReports, CDC}) \\
$\alpha_u$ & $0.1065$ & Fitted (calibrated using US COVID-19 data \cite{dong2020interactive, Dong_tool, WHO_SituationReports, CDC}) \\ \hline 
\end{tabular}
\label{ParVals1}
\end{table}

\begin{figure}[hbtp]
 \centering
 \includegraphics[width=0.8\textwidth]{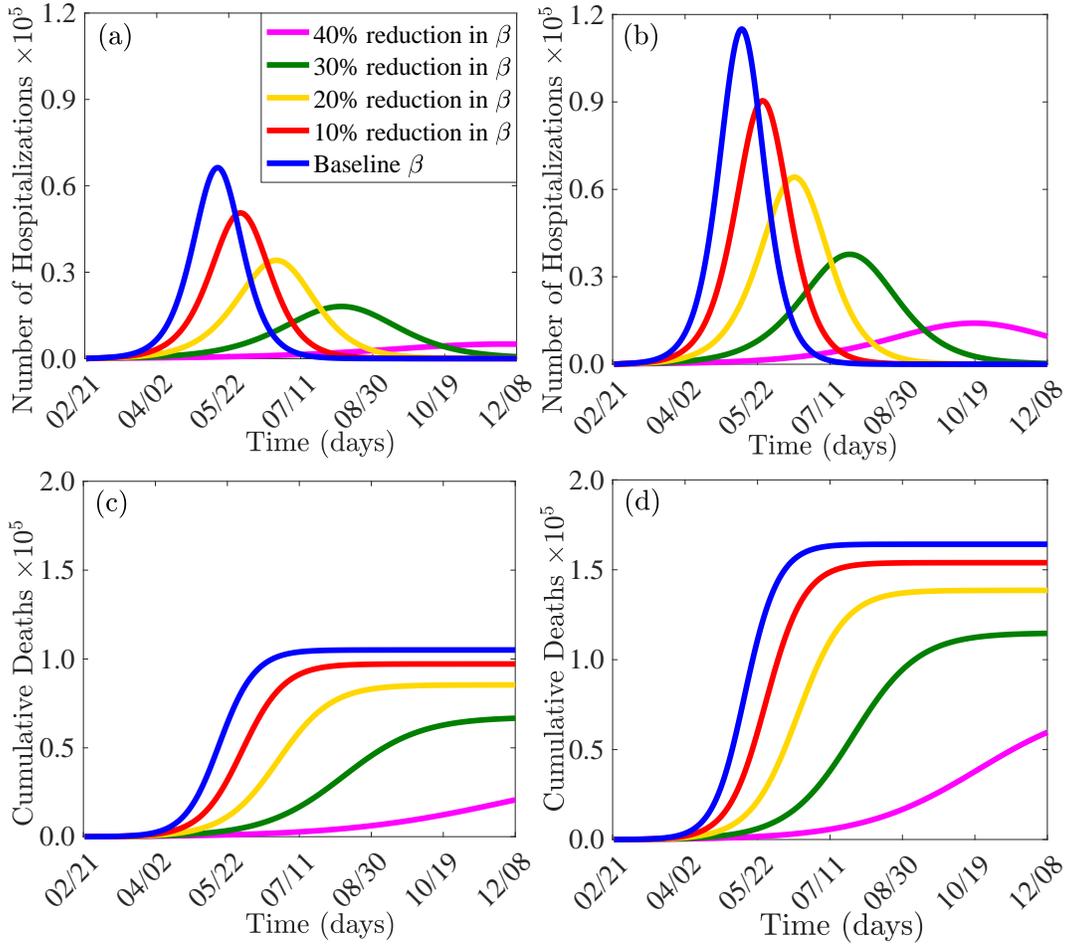}
 \caption{Effect of social-distancing ($\beta$).  Simulations of the model $\eqref{model}$, showing daily hospitalizations (and self-isolation) and cumulative mortality, as a function of time, for various values of social-distancing effectiveness (measured in terms of efficacy and adherence/coverage levels). (a) Daily hospitalizations for the state of New York (b) Daily hospitalizations in the entire US (c) Cumulative mortality for the state of New York (d) Cumulative mortality for the entire US. Parameter values used are as given in Tables \ref{ParVals} and \ref{ParVals1}, with various values of $\beta$.}
 \label{NY_US_SocDis_cumulativedeaths}
 \end{figure}

We simulated the model to, first of all, assess the impact of social-distancing (which, in our study, extends beyond individuals staying 2 meters (or 6 feet) apart to include school and non-essential business closures, staying at home, avoiding large gatherings, etc).  Further, in our study, we measured the effect of social-distancing by the overall reduction in the baseline value of the community contact rate parameter ($\beta$).  The simulation results obtained, depicted in Figure \ref{NY_US_SocDis_cumulativedeaths}, show a projected  $66,300$ patients in hospital (or in self-isolation) at the pandemic peak, expected to be attained on May 5, 2020 (Figure \ref{NY_US_SocDis_cumulativedeaths} (a)) and  $105,100$ cumulative number of deaths (Figure \ref{NY_US_SocDis_cumulativedeaths} (c)) for the state of New York under the baseline scenario (i.e., for the baseline level of social-distancing). Similarly, the projections for the entire US, under the baseline nation-wide social-distancing scenario, are $115,000$ daily hospitalizations at the pandemic peak (Figure \ref{NY_US_SocDis_cumulativedeaths} (b)) and $164,000$ cumulative number of deaths (Figure \ref{NY_US_SocDis_cumulativedeaths} (d)). It is noteworthy that our projection for the cumulative mortality for the entire US (of $164,000$) falls markedly below the 2.2 million mortality projected by Ferguson {\it et al.} \cite{ferguson2020impact}. Our US-wide mortality projection, however, falls within the range of ($38,243, 162,106$) estimated  by Murray {\it et al.} \cite{covid2020forecasting}. When social-distancing is improved above the baseline effectiveness levels (i.e., increase in efficacy and adherence/coverage of social-distancing), Figure \ref{NY_US_SocDis_cumulativedeaths} shows a dramatic reduction of COVID-19 burden for both New York state and the entire US. In particular, for a social-distancing regimen that reduces the contact rate parameter $\beta$ by $10\%$ from its baseline value, the expected number of daily hospitalizations/isolation of confirmed cases at the peak of the pandemic decreases to $50,380$ (corresponding to a $24\%$ decrease in hospitalizations/isolation from baseline) for the state of New York.  Similarly, nation-wide hospitalizations/isolation of confirmed cases at the peak of the pandemic decreases by $21\%$ to $89,930$. Furthermore, for a highly-effective social-distancing strategy (such as a social-distancing strategy that results in at least $40\%$ reduction in the baseline value of $\beta$), the peak hospitalizations/isolation of confirmed cases for New York state and the entire US dramatically reduce to $5,000$ and $14,000$, respectively (this represents a $92\%$ and 88\% reduction in peak hospitalizations for the state of New York and nationwide, respectively). Similarly, for this scenario, the cumulative mortality for New York state and the entire US reduce, respectively, to $20,700$ and $59,600$. Thus, implementing a highly-effective social-distancing strategy (which can reduce the baseline community contact rate, $\beta$, by at least $40\%$) will avert over $80\%$ and $64\%$ of the predicted baseline deaths in New York state and nationwide, respectively. The effectiveness levels and coverage of social-distancing in New York state and the entire US has greatly improved, by April 2, 2020 \cite{Yang, Layne, Levenson, Begley}, to the extent that it is plausible to assume that $40\%$ reduction in the baseline value of $\beta$ has already been achieved in both the state of New York and nationwide.  Therefore, this study shows that the state of New York and the entire US could have recorded catastrophic COVID-19-induced mortality (between $100,000$ to $200,000$) if not for the high effectiveness levels and coverage of the strict social-distancing measures implemented in the state and nationwide.  Our study suggests that if the current level of social-distancing effectiveness and coverage is maintained through May or June 2020, in the state of New York and nationwide, COVID-19 can be eliminated from both the state and the entire nation.  Extending the simulations for Figure \ref{NY_US_SocDis_cumulativedeaths} shows that the current level of the social-distancing regimen in the state of New York should be extended until late September 2021 to guarantee the elimination of COVID-19 (in the context of Figure \ref{NY_US_SocDis_cumulativedeaths}, COVID-19 elimination is measured in terms of when the cumulative mortality stabilizes).  Similarly, for the entire US, social-distancing needs to be maintained until March 2021.
 
\par
\begin{figure}[hbtp]
 \centering
 \includegraphics[width=0.9\textwidth]{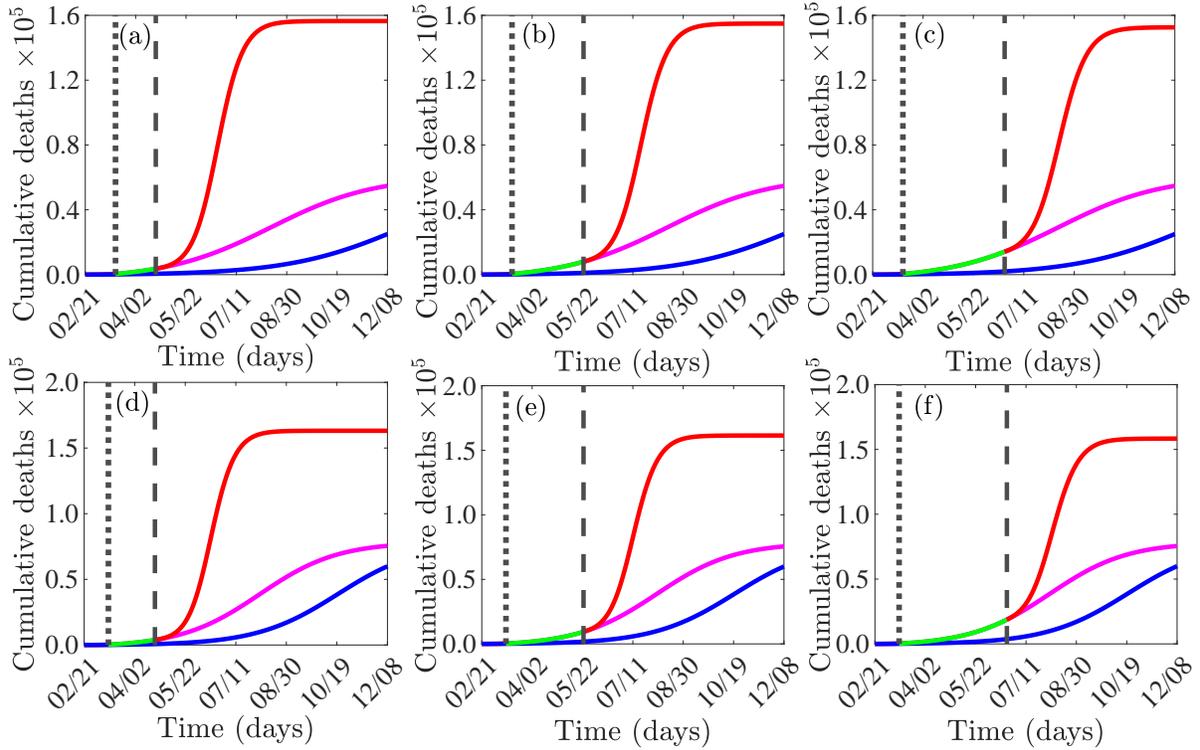}
 \caption{Effect of duration and timing of the termination of the strict social-distancing ($\beta$) measures currently in place in New York state and the entire US. Simulations of the model $\eqref{model}$, showing the effect of the duration and timing of the termination of the strict social-distancing measures against COVID-19. (a) Cumulative mortality for the state of New York if the current strict social-distancing regimen is terminated by the end of April 2020.  (b) Cumulative mortality for the state of New York if the current strict social-distancing regimen is terminated by the end of May 2020. (c) Cumulative mortality for the state of New York if the current strict social-distancing regimen is terminated by the end of June 2020.
 (d) Cumulative mortality for the entire US if the current strict social-distancing regimen is terminated by the end of April 2020.  (e) Cumulative mortality for the entire US if current strict social-distancing regimen is terminated by the end of May 2020. (f) Cumulative mortality for the entire US if current strict social-distancing regimen is terminated by the end of June 2020. Blue curves represent the effect of implementing the current strict social-distancing measures from the very beginning of the COVID-19 pandemic, in New York state (March 1, 2020) and the entire US (January 20, 2020), and maintained until early December, 2020. Magenta curves represent the effect of implementing the current strict social-distancing measures on the actual days they were implemented in New York state (March 22, 2020) and the entire US (March 16, 2020) and maintained until early December, 2020. Red curves represent the effect of terminating the current strict social-distancing regimen by the end of April, 2020 (Plots (a) and (d)), end of May, 2020 (Plots (b) and (e)) and end of June, 2020 (Plots (c) and (f)). With the exception of blue curves, solid dotted vertical lines indicate the start of the current strict social-distancing regimen (March 16, nationwide, and March 22, 2020 for the state of New York), dashed vertical lines indicate when the current strict social-distancing is terminated, and the green horizontal line segments indicate the duration of the strict social-distancing regimen. Parameter values used are as given in Tables \ref{ParVals} and \ref{ParVals1}, with various values of $\beta$.}
 \label{NY_US_SocDis_Timing_cumulativedeaths}
 \end{figure}

 Additional simulations were carried out to assess the population-level impact of the duration and timing of when to terminate the current strict social-distancing protocols. For the best-case scenario, where the current strict social-distancing protocols were assumed to be implemented right from the very beginning of the COVID-19 pandemic in New York state (March 1, 2020) and the entire US (January 20, 2020) and maintained until early December, 2020, the results obtained for the cumulative mortality recorded for New York state and the entire US are $25,000$ and $60,000$, respectively.   This represents $76\%$ and $63\%$ reductions, respectively, in the cumulative mortality for New York state and the entire US, in comparison to the baseline scenario (i.e., worst-case scenario where the social-distancing and other community contacts-reduction strategies have not been implemented at the stringent levels) (blue curves in Figures \ref{NY_US_SocDis_Timing_cumulativedeaths} (a)-(f)). Furthermore, if the social-distancing regimens were implemented on the days they were officially implemented in New York state (March 22, 2020) and the entire US (March 16, 2020), but maintained until early December, 2020, the cumulative mortality to be recorded will be $55,000$ and $75,500$.  This represents  
$48\%$ and $53\%$ reductions, respectively, in the cumulative mortality from the baseline (magenta curves in Figures \ref{NY_US_SocDis_Timing_cumulativedeaths} (a)-(f)).  

\par The effect of the timing of when to terminate the current strict social-distancing protocols was also monitored. Our simulations show that terminating the current strict social-distancing by the end of April 2020 (i.e., the $40\%$ reduction in the baseline value of $\beta$ is now lost due to the termination of the social-distancing measures), a significant rebound of COVID-19 burden will be recorded in as early as July 2020.  In particular, New York state will record $144,000$ deaths representing a $37\%$ increase from the baseline scenario (Figure \ref{NY_US_SocDis_Timing_cumulativedeaths} (a)), while the entire US will record up to $156,000$ deaths. This represents a mere $5\%$ reduction of cumulative mortality, in relation to the baseline scenario (Figure \ref{NY_US_SocDis_Timing_cumulativedeaths} (d)). In other words, the early termination of the current strict social-distancing measures (by the end of April 2020) will result in catastrophic COVID-19 burden, similar to the dire projections made for the pre-social-distancing period (i.e., all the gains of the social-distancing and other control and mitigation measures will essentially be lost). However, if the strict-distancing measures were to be terminated by the end of May, 2020, the cumulative mortality figures are projected to be $91,800$ for New York state and $118,300$ for the entire US. This represents a $13\%$ and $28\%$ reduction, respectively, in the baseline cumulative mortality (Figures \ref{NY_US_SocDis_Timing_cumulativedeaths} (b) and (e)).  Finally, if the social-distancing measures are terminated at the end of June, 2020, the projection for the cumulative mortality figures are $33,200$ for New York state and $50,300$ for the entire US. This represents 68\% and 69\% reductions, respectively, in the baseline cumulative mortality (Figures \ref{NY_US_SocDis_Timing_cumulativedeaths} (c) and (f)). These projected mortality numbers, for the early termination of social-distancing, fall within the range given by Murphy {\it et al.} \cite{covid2020forecasting}. Our study clearly shows that the clamor to relax or terminate the social-distancing measures (that have proven to be hugely successful in both the state of New York and the entire US), as part of the move to re-open the state and the country, would undoubtedly trigger a devastating rebound of COVID-19 in both New York state and the entire US. Data has already shown that certain countries that have relaxed the successfully-implemented social-distancing measures, such as Taiwan, Hong Kong, and South Korea, are now witnessing a rebound of COVID-19 \cite{Wired}.  In particular, Hong Kong announced 84 newly-confirmed cases on March 28 (followed by over $70$ new daily cases in the next three days). Further, Taiwan reported more than $20$ newly-confirmed cases {\it per} day in mid-March (up from barely $5$ cases {\it per} day late in January).  South Korea reported $83$ newly-confirmed cases on April 3, 2020 \cite{Wired}. 

\par It is worth mentioning that the aforementioned simulations for the effect of social-distancing were carried out for the case where other interventions (contact-tracing, quarantine, face-mask, usage etc.) are also implemented (at their baseline values in Tables \ref{ParVals} and \ref{ParVals1}). If face-masks are not used, then the above cumulative numbers will be even more catastrophic.  For instance, if the strict social-distancing protocol is terminated in New York state by April 30, 2020, and no face-mask-based intervention is implemented, about $150,800$ deaths will be recorded by July 2020.  Furthermore, terminating the social-distancing protocols by end of May 2020 or end of June 2020 will result in projected $108,500$ and $44,300$ deaths, respectively, in the state. The corresponding numbers for the entire US (for the case where mask-based intervention is not implemented) are projected to be $167,000, 148,000$, and $91,900$, respectively. Thus, this study strongly suggest that utmost caution should be exercised before terminating the current strict social-distancing protocols being implemented in the state of New York and nationwide.  At the very least, a careful state-by-state (or county-by-county) phase withdrawal (based on the updated COVID-19 incidence, mass testing data, and proximity to COVID-19 hot spots) should be carried out.

\par
\begin{figure}[H]
 \centering
 \includegraphics[width=0.8\textwidth]{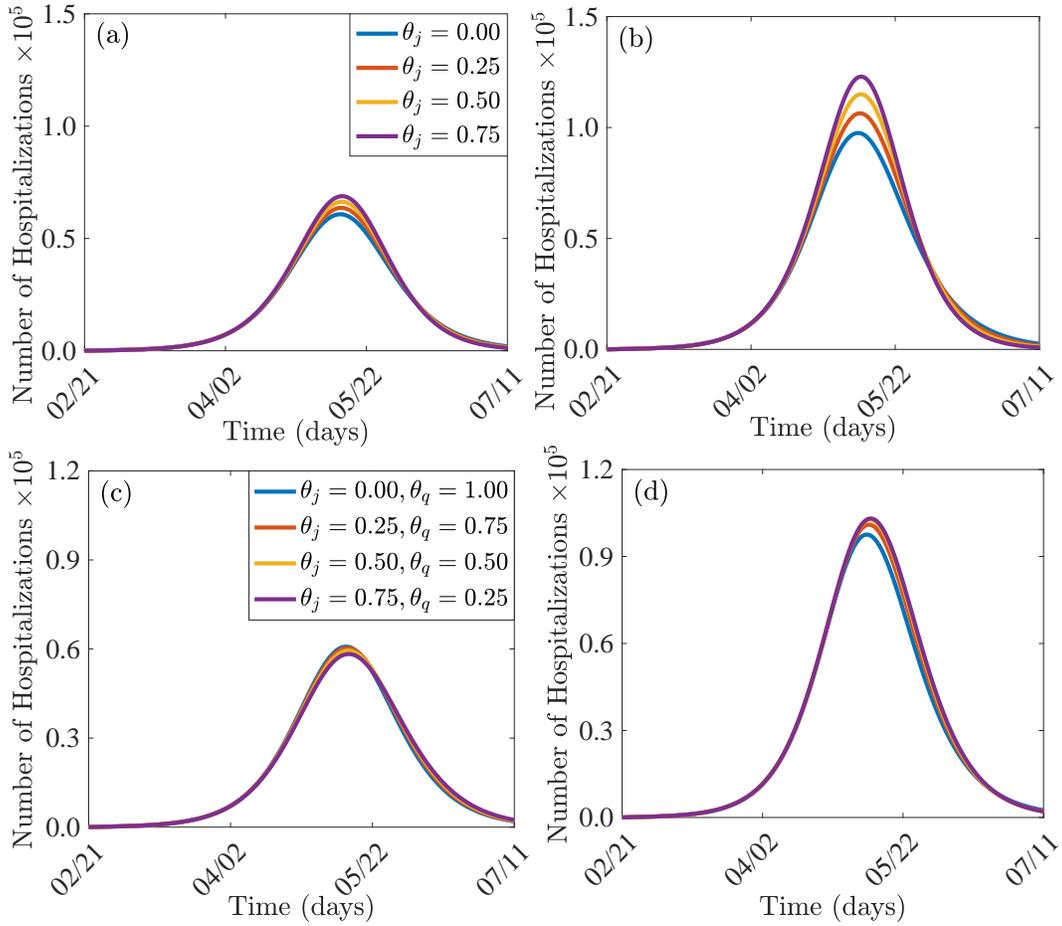}
\caption{Effect of quarantine of individuals suspected of being exposed to COVID-19.  Simulations of the model $\eqref{model}$, showing daily hospitalizations, as a function of time, for various values of the efficacy of quarantine ($\theta_j$) to prevent infection during quarantine and efficacy of isolation or hospitalization to prevent isolated/hospitalized patients from mixing with the rest of the population ($\theta_q$).  (a) Effect of quarantine efficacy ($\theta_j$) on daily hospitalizations for the state of New York (b) Effect of quarantine efficacy ($\theta_j$) on daily hospitalizations for the the entire US. (c) Effect of quarantine ($\theta_j$) and isolation ($\theta_q$) on daily hospitalizations for the state of New York.  (d) Effect of quarantine ($\theta_j$) and isolation ($\theta_q$) on daily hospitalizations for the entire US. Parameter values used are as given in  Tables \ref{ParVals} and \ref{ParVals1}, with various values of $\theta_j$ and $\theta_q$.}
\label{quarantine}
\end{figure}

 The effect of quarantine of individuals suspected of being exposed to COVID-19 is monitored by simulating the model \eqref{model} using the baseline parameter values and various levels of effectiveness of quarantine to prevent the acquisition of infection during quarantine ($\theta_j$). The results obtained, depicted in Figure \ref{quarantine}, show that quarantine of susceptible individuals has only marginal impact in reducing COVID-related hospitalizations for both New York state (Figure \ref{quarantine} (a)) and the entire US (Figure \ref{quarantine} (b)). In particular, at the baseline quarantine efficacy ($\theta_j=0.5$), the state of New York will record $66,000$ daily hospitalizations. The implementation of a perfect quarantine in the state (i.e., $\theta_j=0$) reduces the number of hospitalizations marginally to $60,000$.  The numbers for the entire US for the baseline and perfect quarantine are $115,000$ and $97,000$, respectively.  The marginal effect of quarantine in minimizing COVID-related hospitalizations is even more pronounced when the isolation strategy is perfect (Figures \ref{quarantine} (c) and (d)). That is, for the epidemiological scenario where the isolation (or hospitalization) of confirmed cases is perfect (that is, individuals in isolation at home or in hospital are not part of the actively-mixing population, so that $\theta_q = 1$), the community-wide implementation of mass quarantine of individuals suspected of being exposed to COVID-19 will have very marginal impact on COVID-19 burden (measured in terms of reductions in COVID-19 hospitalizations). This result is consistent with that reported in \cite{day2006}. Thus, this study suggests that, since self-isolation and isolation in hospitals have been implemented at high effectiveness levels in both the state of New York and in the entire US, the mass quarantine of suspected cases may not be a cost-effective public health strategy for combating the spread of COVID-19 in both New York state and the entire US. Figures \ref{quarantine} (c) and (d) illustrate the dynamics of the model \eqref{model} for various effectiveness levels of quarantine and isolation, for both New York state and the US, further emphasizing the marginal nature of quarantine (even for the cases where isolation was not implemented at a perfect level) in minimizing COVID-19 hospitalizations.  

\par
\begin{figure}[H]
\centering
 \includegraphics[width=0.9\textwidth]{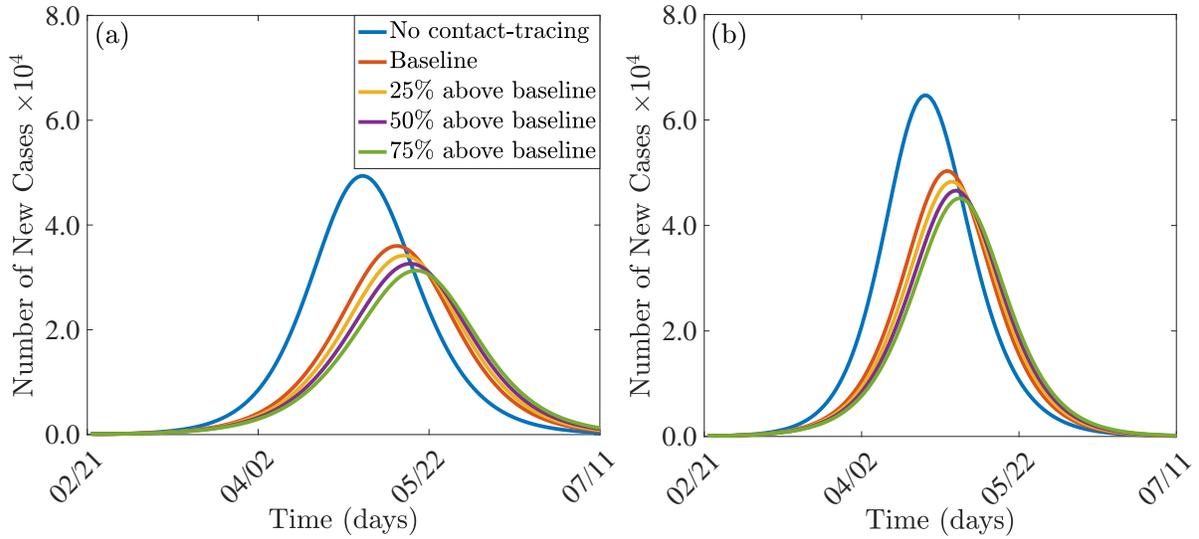}
\caption{Effect of contact-tracing ($\alpha_u$ and $\sigma_a$).  Simulations of the model $\eqref{model}$, showing the number of new COVID-19 cases for various levels of contact-tracing effectiveness (measured based on increases in the values of the contact-tracing parameters, $\alpha_u$ and $\sigma_a$ from their baseline values, as a function of time. (a) Number of new cases for the state of New York (b) Number of new cases for the entire US. Parameter values used are as given in Tables \ref{ParVals} and \ref{ParVals1}, with various values of $\alpha_u$ and $\sigma_a$.}
\label{fig:contact-tracing}
\end{figure}

 The effect of contact-tracing (measured in terms of the detection of asymptomatic cases, following testing/ diagnosis of a confirmed COVID-19 case they may have had close contacts with or random testing) on the transmission dynamics and control of the COVID-19 pandemic is also monitored by simulating the model \eqref{model} using the baseline parameter values in Table \ref{ParVals} and various values of the contact-tracing parameters ($\alpha_u$ and $\sigma_a$). In particular, the simulations are run by increasing the values of $\alpha_u$ and $\sigma_a$, simultaneously (and by the same amount) from their respective baseline values. Figure \ref{fig:contact-tracing} depicts the solution profiles obtained, showing the worst case scenario of $49,400$ cases in the state of New York and $64,600$ cases nationwide on the day the pandemic peaks (on April 26, 2020) if no contact tracing is implemented. If implemented at its baseline rate, contact tracing reduces the size of the pandemic peak number of new COVID-19 cases by $27\%$ for the state of New York, and by $22\%$ nationwide, while a $75\%$ improvement in contact-tracing will reduce the predicted number of confirmed cases to approximately $31,300$ for the state of New York and $41,200$ nationwide. This represents $13\%$ and $10\%$ reduction from baseline and shows that while contact-tracing implemented even only at baseline is important in reducing the size of the pandemic peak number of new COVID-19 cases, investing much resources towards contact-tracing beyond the baseline rate might not be cost-effective.

\par
\begin{figure}[H]
\centering
 \includegraphics[width=1.0\textwidth]{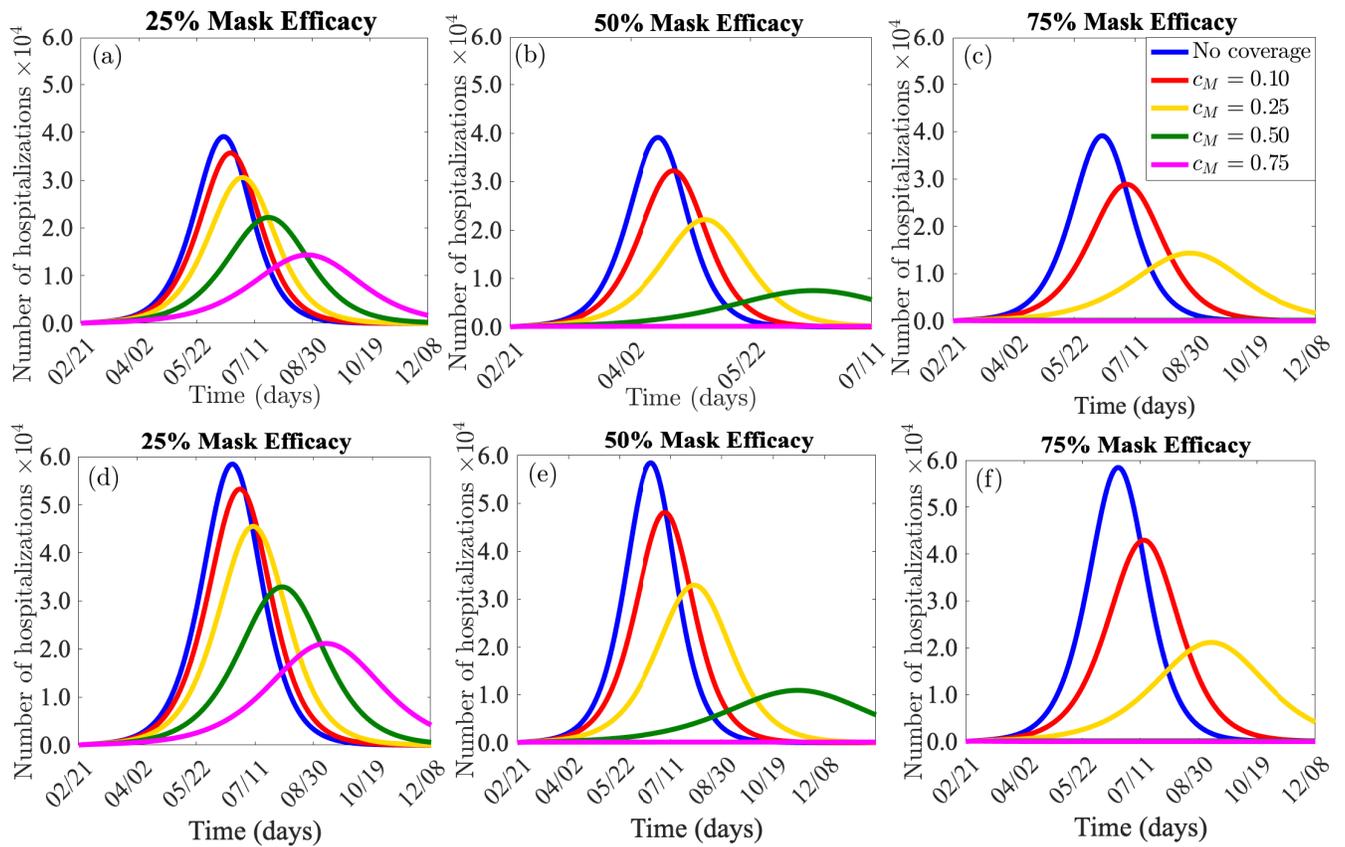}
\caption{Effect of face-masks use in public.  Simulations of the model $\eqref{model}$, showing daily hospitalizations, as a function of time, for various efficacies of of face-masks ($\epsilon_M$) and coverage ($c_M$). (a) $25\%$ mask efficacy for the state of New York.  (b) $50\%$ mask efficacy for the state of New York. (c) $75\%$ mask efficacy for the state of New York.  (d) $25\%$ mask efficacy for the entire US.  (d) $50\%$ mask efficacy for the entire US. (c) $75\%$ mask efficacy for the entire US. Parameter values used are as given in Tables \ref{ParVals} and \ref{ParVals1}, with various values of $\epsilon_M$ and $c_M$.}
\label{mask_efficacy}
\end{figure}

\begin{figure}[H]
\centering
 \includegraphics[width=0.8\textwidth]{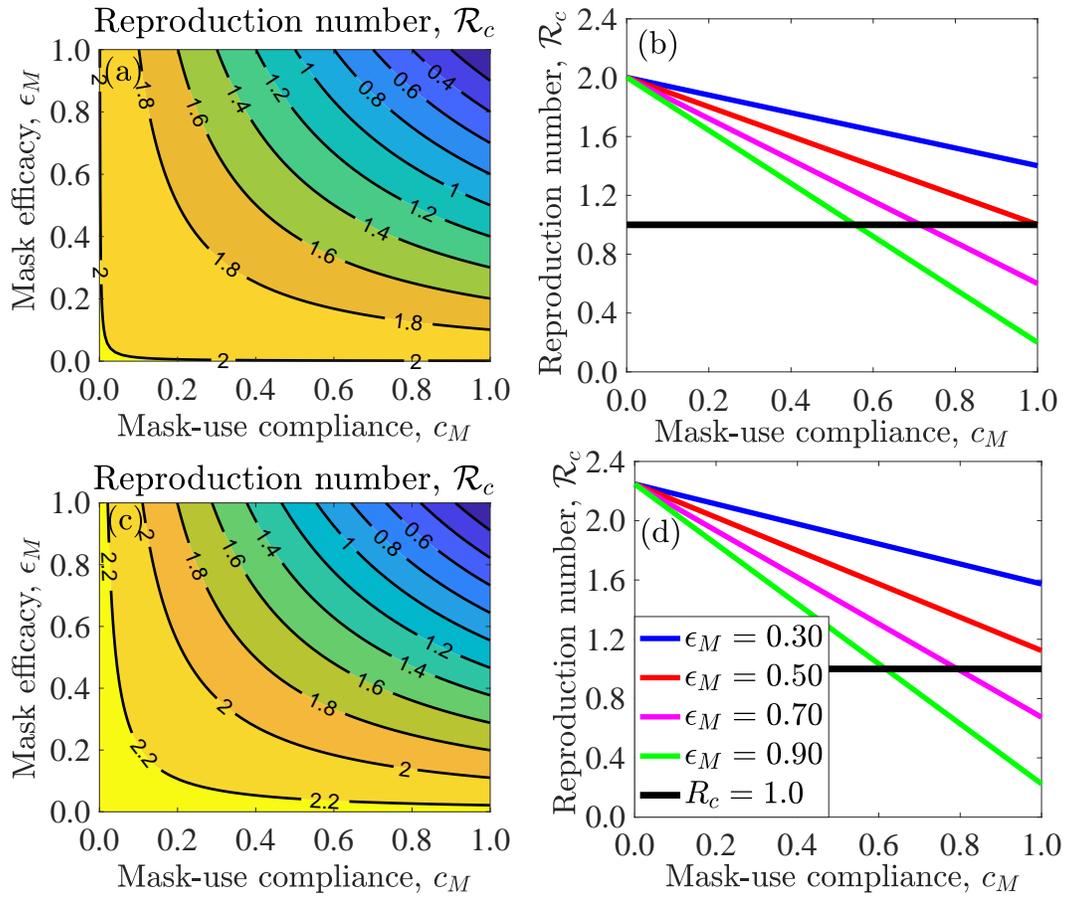}
 \caption{Effect of face-masks use in public. (a) Contour plot of the control reproduction number ($\mathscr{R}_c$), as a function of face-mask efficacy ($\epsilon_M$) and mask coverage ($c_M$), for the state of New York. (b) Profile of the control reproduction number (${\mathscr R}_c$), as a function of face-mask coverage ($c_M$) for the state of New York. (c) Contour plot of of the control reproduction number ($\mathscr{ R}_c$), as a function of face-mask efficacy ($\epsilon_M$) and mask coverage ($c_M$), for the entire US. (d) Profile of the control reproduction number (${\mathscr R}_c$), as a function of mask coverage ($c_M$) for the entire US.} Parameter values used are as given in Tables \ref{ParVals} and \ref{ParVals1}, with various values of $\epsilon_M$ and $c_M$.
 \label{contour}
 \end{figure}

 Simulations were further carried out to assess the population-level impact of the widespread use of masks in public, by running the model \eqref{model} with various values of mask efficacy ($\epsilon_M$) and coverage ($c_M$). The results obtained, depicted in Figure \ref{mask_efficacy}, show a marked decrease in the number of hospitalizations, for both New York state (Figures \ref{mask_efficacy} (a), (b) and (c)) and the entire US (Figures \ref{mask_efficacy} (d), (e) and (f)), with increasing values of the mask efficacy and coverage. Further, using an efficacious mask, such as a  mask of efficacy 50\%, can greatly flatten the pandemic curve, in addition to significantly reducing the burden of the pandemic (measured, in this case, in terms of hospitalizations).  However, such a mask will fail to lead to the elimination of the disease (Figure \ref{mask_efficacy} (b)). It is worth emphasizing that, although the use of masks with low efficacy may not lead to disease elimination, they still are highly useful by causing a significant decrease in the burden of the pandemic (i.e., significantly reduce hospitalizations) if a significant proportion of the populace wear them. For instance, if $75\%$ of the populace in New York or the entire US wear masks with efficacy as low as $25\%$ (i.e., cloths masks), the number of hospitalizations will be reduced by $63\%$ and $64\%$, respectively (compared to the scenario were masks were not used) (Figures \ref{mask_efficacy} (a) and (d)). A contour plot of the reproduction number of the model (${\mathscr R}_c$), as a function of masks efficacy ($\epsilon_M)$ and compliance ($c_M$) is depicted in Figure \ref{contour}. If masks of higher efficacy, such as surgical masks (with estimated efficacy $\ge 70\%$) are used in the state of New York, disease elimination is, indeed, feasible if at least 70\% of the populace wear the masks (Figures \ref{contour} (a) and (b)). Similar results were obtained for the entire US (Figures \ref{contour} (c) and (d)).

\begin{figure}[H]
\centering
 \includegraphics[width=0.8\textwidth]{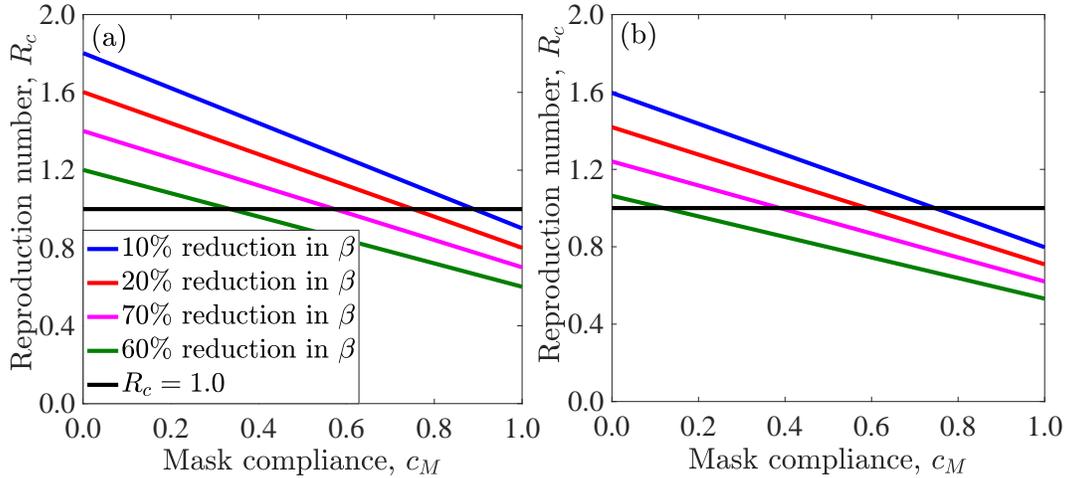}
 \caption{Effect of combined use of face-masks in public and strict social-distancing. Profile of the control reproduction number} (${\mathscr R}_c$), as a function of mask coverage ($c_M$) for different percentage reductions in the baseline value of the effective contact rate ($\beta$) for (a) the state of New York. (b) the entire US. Parameter values used are as given in Tables \ref{ParVals} and \ref{ParVals1}, with various values of $\epsilon_M$ and $c_M$.
 \label{Mask_SocialDist}
 \end{figure}

Additional simulations were carried out to assess the combined impact of public face-masks use strategy and strict social-distancing strategy (which reduces the baseline value of the community transmission parameter, $\beta$, by $40\%$) on the control of COVID-19 in New York state and the entire US. The results obtained, depicted in Figure \ref{Mask_SocialDist}, show that, combining the strict social-distancing strategy with a strategy based on using moderately-effective face-masks (with efficacy $\epsilon_M \ge 0.5)$ in public, will lead to the elimination of the disease in New York state if only $30\%$ of the population use face-masks in public (Figure \ref{Mask_SocialDist} (a)). This clearly shows that disease elimination in New York state is more feasible if the face-masks-based strategy is combined with the strict social-distancing strategy. Similar results were obtained for the entire US (Figure \ref{Mask_SocialDist} (b)), where, in this case, only $10\%$ compliance in mask usage in public will be needed for COVID-19 elimination.

In summary, the above simulations show that the use of face-masks (even those with low efficacy, but with high coverage) in public offers significant community-wide impact in reducing and mitigating the burden of COVID-19 in both New York state and the entire US.  In other words, the use of low efficacy face masks with high coverage is always useful.  Further, combining the face-masks use strategy with a strategy based on the implementation of strict social-distancing is more effective in curtailing (and eliminating) COVID-19, in comparison to the singular implementation of either strategy.

\section{Discussion and Conclusions}
The world is currently experiencing a devastating pandemic of a novel Coronavirus (caused by SARS-CoV2) that emerged in Wuhan city of China in December of 2019. The deadly COVID-19 pandemic has spread to over 210 countries, causing over 1.9 million cases and 125,000 deaths worldwide (with some parts of Asia, Europe and, now, the US suffering the brunt of the burden). There is currently no safe and effective vaccine for use in humans against COVID-19. There is also no safe and effective antiviral.  Consequently, control and mitigation efforts against COVID-19 are limited to non-pharmaceutical interventions, such as social-distancing (which involves keeping a physical distance of at least 6 feet from other humans in public, lockdowns of communities, closure of schools, malls, places of worships and other gathering places), quarantine of suspected cases, contact-tracing, isolation (at home or in hospital) of confirmed cases and the use of face masks (both low quality cloth masks and the higher quality surgical masks) in public. This study is based on the design, analysis and simulations of a new mathematical model for providing deeper insights into the transmission dynamics and control of COVID-19 in a community. Specifically, the model designed in this study was parametrized using COVID-19 data from the US state of New York and the entire US population. The model was used to assess the population-impact of the aforementioned control and mitigation interventions.

We parameterized the model using COVID-19 data from New York state and the entire US, and extensive numerical simulations were carried out using the parametrized model to assess the population-level impact of the various intervention strategies. With the baseline levels of the four main intervention strategies considered (social-distancing, quarantine/isolation, contact-tracing and the use of face-masks), the state of New York is projected to see a peak of the pandemic around mid April, 2020 (with $66,300$ number of hospitalizations/isolation of confirmed cases and $105,100$ deaths at the peak), while the entire US will see its peak around end of April, 2020 (with $115,000$ hospitalizations/isolation of confirmed cases and $164,000$ deaths at the peak). Our projections for baseline (worst-case) mortality for the US are much lower than the 2.2 million deaths suggested by Ferguson {\it et al.} in the absence of interventions \cite{ferguson2020impact}, but fall within the range estimated in \cite{covid2020forecasting}. Our projected numbers for COVID-19 burden (morbidity and mortality) dramatically decreases if strict social-distancing measures are implemented at high adherence levels. For instance, it was shown that strict compliance to the statewide lock-down in New York state (which corresponds to reducing the baseline contact rate in our model by at least $40\%$) will reduce the peak values for hospitalizations/isolation of confirmed cases and mortality by $92\%$ and $80\%$, respectively.  Similarly, the peak values for cases and mortality in the entire US (if a nation-wide lockdown capable of reducing baseline contact rate by at least $40\%$) will decrease by $88\%$ and $64\%$, respectively.

The duration and timing of the termination of strict social-distancing measures are critically-important in the battle to effectively combat pandemics of respiratory diseases, such as the devastating COVID-19 pandemic. While the rapid implementation of strict social-distancing measures (during the early stage of the pandemic), maintained over a relatively long period of time (e.g., until the summer), will undoubtedly effectively combat the burden of the pandemic. Early termination of these measures will cause catastrophic outcomes.  For instance, our study shows that relaxing or terminating the strict social-distancing measures in the state of New York and the US as a whole by end of April 2020 will trigger a devastating second wave, generating COVID-19 burden similar to those obtained during the pre-strict-social-distancing time in the state and in the entire nation by the end of July 2020 (with cumulative mortality numbers in the range reported in \cite{covid2020forecasting} for both New York state and the entire US). In particular, up to $144,000$ and $156,000$ cumulative deaths will have been recorded in the state of New York and the entire US if the social-distancing measures are shut down by the end of April, 2020. Extending the termination of social-distancing, such as to end of June 2020, significantly reduces the likelihood of a second wave (in addition to significantly reducing the associated burden of the pandemic). Consequently, a great deal of caution must be exercised before decisions are made to relax or terminate the existing highly-successful social-distancing protocols in both the state of New York and the entire US.  It is noteworthy that countries that have recently relaxed these measures, such as Taiwan, Hong Kong and South Korea, have already started seeing a rebound of COVID-19. Our study suggests that the decision to relax or terminate the social-distancing measures should by on a case-by-case (i.e., state-by-state or county-by-county) basis, and should be informed by updated COVID-19 incidence and mortality data, number of COVID-19 tests (both antibody and surveillance tests) and proximity of a locality to COVID-19 hot spots. In particular, our study shows that strict social-distancing should be maintained until the year 2021  (up to late September 2021 for New York state, and early March, 2021 for the entire US) to eliminate COVID-19.

\par Quarantine of people suspected of being exposed to a respiratory disease is perhaps the oldest public health control measure in human history.  Our study shows that widescale implementation of quarantine intervention may not be very effective (in minimizing the burden of COVID-19) if the strategy of isolating confirmed cases is effective.  In other words, our study suggests that if isolation can be implemented effectively (high efficacy and coverage), then quarantine of people suspected of contracting COVID-19 may not be necessary. This result is consistent with what was reported by Day {\it et al.} \cite{day2006}. Tracing the contacts of confirmed cases (known as contact-tracing) was also shown to only be marginally-effective in minimizing the burden of the pandemic.  In particular, even if contact-tracing is implemented at the highest possible level (represented in our study based on increases in the contact-tracing parameters by $75\%$), the decrease in the burden of the pandemic recorded was only marginal ($13\%$ and $10\%$ for cases in New York state and nationwide, respectively, and $5\%$ and $3\%$ for mortality for New York state and nationwide, respectively).

\par The use of face-masks in public in times of outbreaks of respiratory diseases has a rich history.  Although quite popular in some parts of the world (notably Asia), the use of face-masks in public is somewhat controversial. This was more evident in the US during the COVID-19, leading, ultimately, to the recommendation to use face-masks (home-made cloths masks) in public by the The US Centers for Disease Control and Prevention (CDC) on April 2, 2020. Our study shows that the use of high efficacy masks (such as surgical masks, with estimated efficacy of at least $70\%$) will lead to a dramatic reduction of COVID-19 burden if its coverage is high enough (at least $70\%$). In fact, our study shows that even face-masks of low efficacy (home-made cloths masks) will lead to a dramatic reduction of disease burden ({\it albeit} this will not lead to the elimination of the disease). For example, even face-mask efficacy of $25\%$ can lead to a $63\%$ and $64\%$ reduction in the number of hospitalizations/isolation of confirmed cases at the pandemic peak in New York state and nationwide if $75\%$ of the population wear face-masks in public. These results are consistent with those reported by Eikenberry {\it et al.} \cite{eikenberry2020}. Furthermore, by generating contour plots for the control reproduction number of the model $\eqref{model}$ (${\mathscr R}_c$), as a function of mask efficiency ($\epsilon_M$) and coverage ($c_M$), our study shows that the use of high efficacy masks (such as surgical masks, with estimated efficacy of $\ge 75\%$) will, indeed, lead to the elimination of COVID-19 (in both the state of New York and in the entire US nation) if the coverage is high enough (about 80\%). This study shows that the use of face-masks in public is always useful, and their population-level impact increases will increases efficacy and coverage.   In particular, even the use of low efficacy masks will greatly reduce the burden of the pandemic if the coverage in their usage in the community is high enough.  Furthermore, our study shows that combining the masks-based strategy with the strict social-distancing strategy is more effective than the singular implementation of either strategy.  For instance, our study shows that COVID-19 can be eliminated from the state of New York if the strict social-distancing measures implemented are combined with a face-masks strategy, using a moderately-effective mask (with efficacy of about $50\%$) if only $30\%$ of the residents of the state wear the masks.  The masks use compliance needed to eliminate the disease nationwide, under this scenario with strict social-distancing nationwide, is a mere $10\%$.  

In summary, our study suggests that, like in the case of the other Coronaviruses we have seen in the past (namely SARS and MERS \cite{yin2018mers}),  COVID-19 is a disease that appears to be controllable using basic non-pharmaceutical interventions, particularly social-distancing and the use of face-masks in public (especially when implemented in combinations). The factors that are obviously critically-important to the success of the anti-COVID-19 control efforts are the early implementation (and enhancement of effectiveness) of these intervention measures, and ensuring their high adherence/coverage in the community.

\section*{Acknowledgments} \noindent One of the authors (ABG) acknowledge the support, in part, of the Simons Foundation (Award $\#$585022) and the National Science Foundation (Award $\#$1917512). CNN acknowledges the support of the Simons Foundation (Award \#627346). One of the authors (ABG) acknowledges the useful conversation with Dieter Armbruster on some aspects of public mask usage.
The authors are grateful to the two anonymous reviewers and the Handling Editor for their very constructive comments, which have significantly enhanced the manuscript.

\end{document}